\documentclass[twocolumn,showpacs,prd,aps,amsmath,nofootinbib,amssymb,eqsecnum,superscriptaddress]{revtex4-1}
\usepackage{mathrsfs} 

\usepackage{color}
\usepackage{hyperref}
\usepackage{graphicx}
\usepackage{bm} 
\usepackage{mathrsfs} 
\newcommand{\der}[2]{\frac{\partial #1}{\partial #2}}

\newcommand{\dert}[2]{{\partial #1}/{\partial #2}}

\newcommand{\w}[1]{\bm{#1}}
\newcommand{\vw}[1]{\vec{\w{#1}}}

\newcommand{\dd}{\bm{\mathrm{d}}}

\newcommand{\be}{\begin{equation}}
\newcommand{\ee}{\end{equation}}
\newcommand{\bea}{\begin{eqnarray}}
\newcommand{\eea}{\end{eqnarray}}
\newcommand{\eps}{\epsilon}
\newcommand{\weps}{\w{\eps}}
\newcommand{\M}{\mathscr{M}}
\newcommand{\Hor}{\mathcal{H}}
\newcommand{\VV}{\mathscr{V}}
\newcommand{\Sp}{\mathcal{S}}
\newcommand{\Sps}{\mathscr{S}}
\newcommand{\D}{\mathrm{d}}
\newcommand{\Pp}{\mathscr{P}}
\newcommand{\Lie}{\w{\mathcal{L}}}
\renewcommand{\th}{\theta}

\def\spose#1{\hbox to 0pt{#1\hss}}
\def\lta{\mathrel{\spose{\lower 3pt\hbox{$\mathchar"218$}}
          \raise 2.0pt\hbox{$\mathchar"13C$}}}
\def\gta{\mathrel{\spose{\lower 3pt\hbox{$\mathchar"218$}}
          \raise 2.0pt\hbox{$\mathchar"13E$}}}

\begin{document}

\title[Extracting black-hole rotational energy]{Extracting black-hole rotational energy: The generalized Penrose process}

\date{31 January 2014} 

\author{J.-P. Lasota}\email[]{lasota@iap.fr}
\affiliation{Institut d'Astrophysique de Paris, UMR 7095 CNRS, UPMC Univ Paris 06, 98bis Bd Arago, 75014 Paris, France}
 \affiliation{Nicolaus Copernicus Astronomical Center, ulica Bartycka 18, PL-00-716
               Warszawa, Poland}
\affiliation{Astronomical Observatory, Jagiellonian University,
               ulica Orla 171, 30-244 Krak{\'o}w, Poland} 
              
\author{E. Gourgoulhon}
\affiliation{LUTH, Observatoire de Paris, CNRS, Universit{\'e} Paris Diderot, 5 place
               Jules Janssen, 92190 Meudon, France}
\author{M. Abramowicz}
\affiliation{Department of Physics, University of Gothenburg,
               SE-412-96 G{\"o}teborg, Sweden}
\affiliation{Nicolaus Copernicus Astronomical Center, ulica Bartycka 18, PL-00-716
               Warszawa, Poland}
\affiliation{Institute of Physics, Silesian University in Opava,
                   Bezru{\v c}ovo n{\'a}m. 13, CZ-746-01 Opava,
                   Czech Republic}
\author{A.~Tchekhovskoy}\thanks{Princeton
                  Center for Theoretical Science Fellow and NASA
                  Einstein Fellow}  
\affiliation{Lawrence Berkeley National Laboratory, 1 Cyclotron Road,
  Berkeley, California 94720, USA}
 \affiliation{University of California Berkeley, Berkeley, Berkeley,
  California 94720-3411}
\affiliation{Center for Theoretical Science, Jadwin Hall, Princeton University,
               Princeton, New Jersey 08544, USA}
   \author{R.~Narayan}
\affiliation{Institute for Theory and Computation, Harvard-Smithsonian Center for
               Astrophysics, 60 Garden Street, MS 51, Cambridge, Massachusetts 02138, USA}

\begin{abstract}
In the case involving particles the necessary and sufficient condition for
the Penrose process to extract energy from a rotating black hole is absorption of 
particles with negative energies and angular momenta. No torque at the black-hole horizon occurs. 
In this article we consider the case of arbitrary fields or matter described by an unspecified, general
energy-momentum tensor $T_{\mu \nu}$ and show that the necessary and sufficient condition for extraction 
of a black hole's rotational energy is analogous to that in the mechanical Penrose process: absorption of negative
energy and negative angular momentum. We also show that a necessary condition for the Penrose process to occur is for the Noether current (the conserved energy-momentum density vector) to be spacelike or past directed (timelike or null) on some part of the horizon. In the particle case, our general criterion for the occurrence of a Penrose process  reproduces the standard result. In the case of  relativistic jet-producing ``magnetically arrested disks'' we show
that the negative energy and angular-momentum absorption condition is obeyed when the Blandford-Znajek mechanism is at work, and hence the high energy extraction 
efficiency up to $\sim 300\%$ found in recent numerical simulations of such accretion flows results from
tapping the black hole's rotational energy through the Penrose process. We show how black-hole rotational energy extraction works in this case by describing the Penrose process in terms of the Noether current.
\end{abstract}


\maketitle

\section{Introduction}

\label{section-Introduction}

Relativistic jets are often launched from the vicinity of accreting black holes. They are observed to
be produced in stellar-mass black-hole binary systems and are believed to be the fundamental part
of the gamma-ray burst phenomenon. Powerful relativistic jets are also ejected by accreting supermassive
black holes in some active galactic nuclei. There is no doubt that the jet-launching mechanism is
related to accretion onto black holes, but there
has been no general agreement as to the ultimate source of energy of these spectacular high energy
phenomena. In principle, relativistic jets can be powered either by the black hole gravitational pull
or by its rotation (spin), with large-scale magnetic fields invoked as energy extractors in both cases.
Black-hole rotational energy
extraction due to weakly magnetized accretion was considered by
\citet{1975PhRvD..12.2959R} (see also
\cite{1978PhRvD..17.1518D}). In the context of strongly magnetized jets,
\citet{Blandford-1977} (hereafter BZ) proposed a model of electromagnetic extraction of  black 
hole's rotational energy based on the analogy with the classical 
Faraday disk (unipolar induction) phenomenon. The difficulty with applying this analogy to a rotating black hole was 
a viable identification of the analogue of the Faraday disk in a setup where the surface of the rotating body 
(the black hole's surface) is causally disconnected from the rest of the Universe. It seems now that this 
problem has been clarified and solved (\cite{Komissarov-2006,Komissarov-2009} and references therein).  Another
subject of discussion about the physical meaning of the BZ mechanism was its relation to the
black-hole rotational energy extraction process proposed by \citet{Penrose-1969}, in which an infalling
particle decays into two in the ergoregion, with one of the decay products being absorbed by the
black hole, and the other one reaching infinity, with energy larger than that of the initial, infalling parent particle {(see \cite{WaghD89} for a review)}.
The energy gain in this (``mechanical") Penrose process is explained by the negative (``seen" from infinity) 
energy of the ergoregion-trapped particle absorbed by the black hole. In the BZ mechanism, particle inertia can be
neglected; therefore it clearly is not a mechanical Penrose process. \citet{Komissarov-2009}
argues that the BZ mechanism is an example of an {\sl energy counterflow}, a black-hole extraction phenomenon 
supposed to be more general than the Penrose process.

In the present article we discuss the relation between any mechanism extracting black-hole rotational energy and the mechanical Penrose process
using a general-relativistic, covariant description of the energy fluxes in the metric of a 
stationary and axisymmetric rotating black hole (this framework encompasses the Kerr metric as
the special case of a black hole surrounded by non-self-gravitating matter). In particular, using energy and angular-momentum conservation laws, we prove that {\sl for {\tt any} matter or
field, tapping the black-hole rotational energy is possible if and only if
negative energy and angular momentum are absorbed by the black hole and no torque at the black-hole horizon is necessary (or possible)}. The conditions on energy and angular-momentum fluxes through the horizon
are analogous to those on particle energy and angular momentum in the mechanical Penrose process. 
From these conditions, we deduce a necessary condition for a general (passive) electromagnetic field configuration to allow black-hole energy extraction through the Penrose process. In the case of stationary, axisymmetric, and force-free fields we obtain the well-known condition \cite{Blandford-1977} on the angular speed of the field lines.
We also describe the Penrose process in terms of the Noether current. This description is 
particularly useful in the description of results of numerical simulations.

Finally, we use our generalized Penrose process framework to interpret the results of recent numerical studies of accretion onto black holes by \cite{Tchekhovskoy-2011,Tchekhovskoy-2012,McKinney-2012}, which indicate that the BZ mechanism can tap  the black-hole rotational energy very efficiently
(efficiency $\eta > 100\%$). 
These simulations are based on
large-scale numerical simulations involving a particular state of
accretion around rotating black holes: ``magnetically arrested disks''
(MADs), first in Newtonian gravity \citep[see, e.g.,][]{Igu2003,Narayan-2003}, and later in GR
\citep[e.g.,][]{Tchekhovskoy-2011}, \cite{Tchekhovskoy-2012}).  MADs were
also called ``magnetically choked accretion flows'' (MCAFs) in \cite{McKinney-2012}. We show that the resulting configurations satisfy the Penrose-process conditions for black-hole energy energy extraction. 

Our results agree, in most respects, with those obtained  
by \citet{Komissarov-2009}.
The difference between the two approaches worth noticing, is that we derive our generalized Penrose condition from the
fundamental, and universally accepted, {\it null energy condition}, while Komissarov introduces a new concept of the {\it energy counterflow}.
This difference will be investigated in a future paper.

More than 30 years ago \citet{Carte79}, analyzing the BZ mechanism in a covariant framework obtained several results similar to ours. Using energy and angular-momentum {\sl rates} (integrated fluxes, while we use energy and angular momentum) he showed the necessity of a negative energy absorption rate at the horizon for this mechanism to operate. Strangely, his paper has almost never been cited in the context of the discussion of the Penrose-BZ process. Our treatment is more general than that of Carter since we use a general energy-momentum tensor, {while Carter considered fields that are time periodic
(cf. Sec.~6.4.2 of Ref.~\cite{Carte79}). Moreover}
 we obtain a new condition on a general electromagnetic field configuration 
 {[Eq.~(\ref{e:nec_cond_EM}) below]} and we apply it to interpret recent numerical simulation of relativistic jet production.

In a recent paper \citep{Penna-13} the MAD simulations have been described in the framework of the so-called ``membrane paradigm" \citep{Thorne-86}.
This picture of the interaction of electromagnetic fields with the black-hole surface has the
advantage of using the analogues of the usual electric and magnetic fields in a 3-D flat space. \citet{Penna-13} showed that the results of MAD simulations can be consistently described in the membrane framework.

\section{The mechanical Penrose process}
\label{Section-Penrose-particles}

\citet{Penrose-1969} considered\footnote{See also \cite{Penrose-1971}} a free-falling
particle that enters the ergosphere of a rotating black
hole with energy $E_1 = - \vw{\eta}\cdot\vw{p}_1$, where $\vw{\eta}$
is the Killing vector associated with stationarity [see also
Eq.~(\ref{e:dodt_eta}) below], $\vw{p}_1$ the particle
4-momentum vector and the dot denotes the spacetime scalar product: 
$\vw{\eta}\cdot\vw{p}_1 = \w{g}(\vw{\eta},\vw{p}_1) = g_{\mu\nu}
\eta^\mu p_1^\nu = \eta_\mu p_1^\mu$. 
Here $\w{g}$ is the metric tensor, whose signature is chosen to be $(-,+,+,+)$.
Note that although  $E_1$ is called an \emph{energy}, it
is not the particle's energy measured by any observer since $\vw{\eta}$ is not a unit
vector (i.e. cannot be considered as the 4-velocity of any observer), except in the
asymptotically flat region infinitely far from the black hole. For this reason $E_1$ is often called 
the \emph{energy at infinity}. The virtue of $E_1$ is to remain constant 
along the particle's worldline, as long as the latter is a geodesic, i.e., as long as
the particle is free falling. In the ergoregion, the particle disintegrates 
into two particles with, say,  4-momenta $\vw{p}_2$ and $\vw{p}_*$. 
Their conserved energies are, respectively, $E_2 = - \vw{\eta}\cdot\vw{p}_2$
and $\Delta E_H = - \vw{\eta}\cdot\vw{p}_*$ (the notation $\Delta E_H$ is
for future convenience). The first particle escapes to infinity, 
which implies $E_2 > 0$, while the second one falls into the black hole. 
Since in the ergoregion $\vw{\eta}$ is a spacelike vector (from the very definition 
of an ergoregion), it is possible to have $\Delta E_H<0$ on certain 
geodesics. The falling particle is then 
called a \emph{negative energy particle}, although its energy measured by any
observer, such as for instance a zero-angular-momentum observer (ZAMO), 
remains always positive. 
At the disintegration point, the conservation of 4-momentum 
implies $\vw{p}_1 = \vw{p}_2 + \vw{p}_*$; taking the scalar product 
with $\vw{\eta}$, we deduce that 
$E_1 = E_2 + \Delta E_H$. Then, as a result of $\Delta E_H < 0$, we get $E_2 > E_1$. 
At infinity, where the constants $E_1$ and $E_2$ can be interpreted 
as the energies measured by an inertial observer at rest with respect to the 
black hole (thanks to the asymptotic behavior of $\vw{\eta}$), one has 
clearly some energy gain: the outgoing particle is more energetic than 
the ingoing one. This is the so-called \emph{mechanical Penrose process} of 
energy extraction from a rotating black hole. 
In other words, the sufficient and necessary condition for 
energy extraction from a rotating black hole is
\be
\label{eq:Penr_E}
    \Delta E_H <0.
\ee
From the condition that energy measured locally by a ZAMO must be non-negative
one obtains (see e.g. \cite{Hartle-2003})
\be
\label{eq:Penr_Epos}
\omega_H \Delta J_H\leq \Delta E_H,
\ee
where $\omega_H$ is the angular velocity of the black hole (defined below) and $\Delta J_H$ is the angular-momentum
of the negative-energy particle absorbed by the black hole, defined by
$\Delta J_H = \vw{\xi} \cdot \vw{p}_*$, where $\vw{\xi}$ is the Killing vector associated 
with axisymmetry. 
Without loss of generality, we take $\omega_H \ge0$. 
Equations~(\ref{eq:Penr_E})-(\ref{eq:Penr_Epos}) imply that $\omega_H \neq 0$  and 
\be
\label{eq:Penr_J}
\Delta J_H < 0 .
\ee

It worth stressing that in the mechanical Penrose process, particles move on geodesics along
which (by construction) energy is conserved. Therefore the negative-energy particle must
originate in the ergoregion, the only domain of spacetime where such particle can exist. In
the general case of interacting matter or fields, negative energy at the horizon does not imply negative energy elsewhere.

Soon after Penrose's discovery that rotating black holes may be energy sources, it was suggested 
that the mechanical Penrose process may power relativistic jets observed in quasars. 
However, a careful analysis by \cite{Bardeen-1972, Wald-1974,
Kovetz-1975, Piran-1977} {(see also \cite{WaghD89})}, showed that it is unlikely that negative energy states,
necessary for the Penrose process to work, may be achieved through the particles
disintegration and/or collision inside the ergosphere. This conclusion has been confirmed more recently by
\cite{Bejger-2012,Harada-12,Zaslavskii-12} for high energy particle collisions. The reason is that in the case of
collisions, the particles with positive energies cannot escape because they must have large but
negative radial momenta. Thus, they are captured (together with the negative energy
particles) by the black hole. {Note that for charged particles evolving
in the electromagnetic field of a Kerr-Newman black hole, the efficiency of the 
mechanical Penrose process can be very large \cite{BhatDD85,WaghD89}.}

Attempts to describe the BZ mechanism as a mechanical Penrose process have been unsuccessful (\cite{Komissarov-2009} 
and references therein).
This leaves electromagnetic processes as the only astro\-phy\-si\-cally realistic way to extract
rotational energy from a rotating black hole.

\section{General relativistic preliminaries}
\label{Section-general-relativistic}

\subsection{The spacetime symmetries}
\label{sub-section-Kerr-symmetries}

The spacetime is modeled by a four-dimensional smooth manifold $\M$ equipped with 
a metric $\w{g}$ of signature $(-,+,+,+)$. 
We are considering a rotating uncharged black hole that is stationary and axisymmetric. 
If the black hole is isolated, i.e., not surrounded by self-gravitating matter or electromagnetic
fields, the spacetime $(\M,\w{g})$ is described by the Kerr metric (see
\ref{ap:Kerr}). 
Here and in Secs. \ref{s:conservation_laws} to \ref{sect:em}, we do not restrict to this case and consider a generic stationary and 
axisymmetric metric $\w{g}$. As already mentioned in Sec.~\ref{Section-Penrose-particles}, we
denote by $\vw{\eta}$ the Killing vector associated with stationarity and by $\vw{\xi}$
that associated with axisymmetry.
In a coordinate system $(x^\alpha)=(t,x^1,x^2,x^3)$ adapted to stationarity, 
i.e. such that 
\be \label{e:dodt_eta}
     \der{}{t} = \vw{\eta} ,
\ee
the components $g_{\alpha\beta}$ of the metric tensor are independent of the coordinate $t$. 
In a similar way,
if the coordinate $x^3$, say, corresponds to the axial symmetry, the components
$g_{\alpha\beta}$ will be independent of this coordinate.

\subsection{The black-hole horizon}
\label{sub-section-Kerr-horizon}

\begin{figure}
\centerline{\includegraphics[height=0.27\textheight]{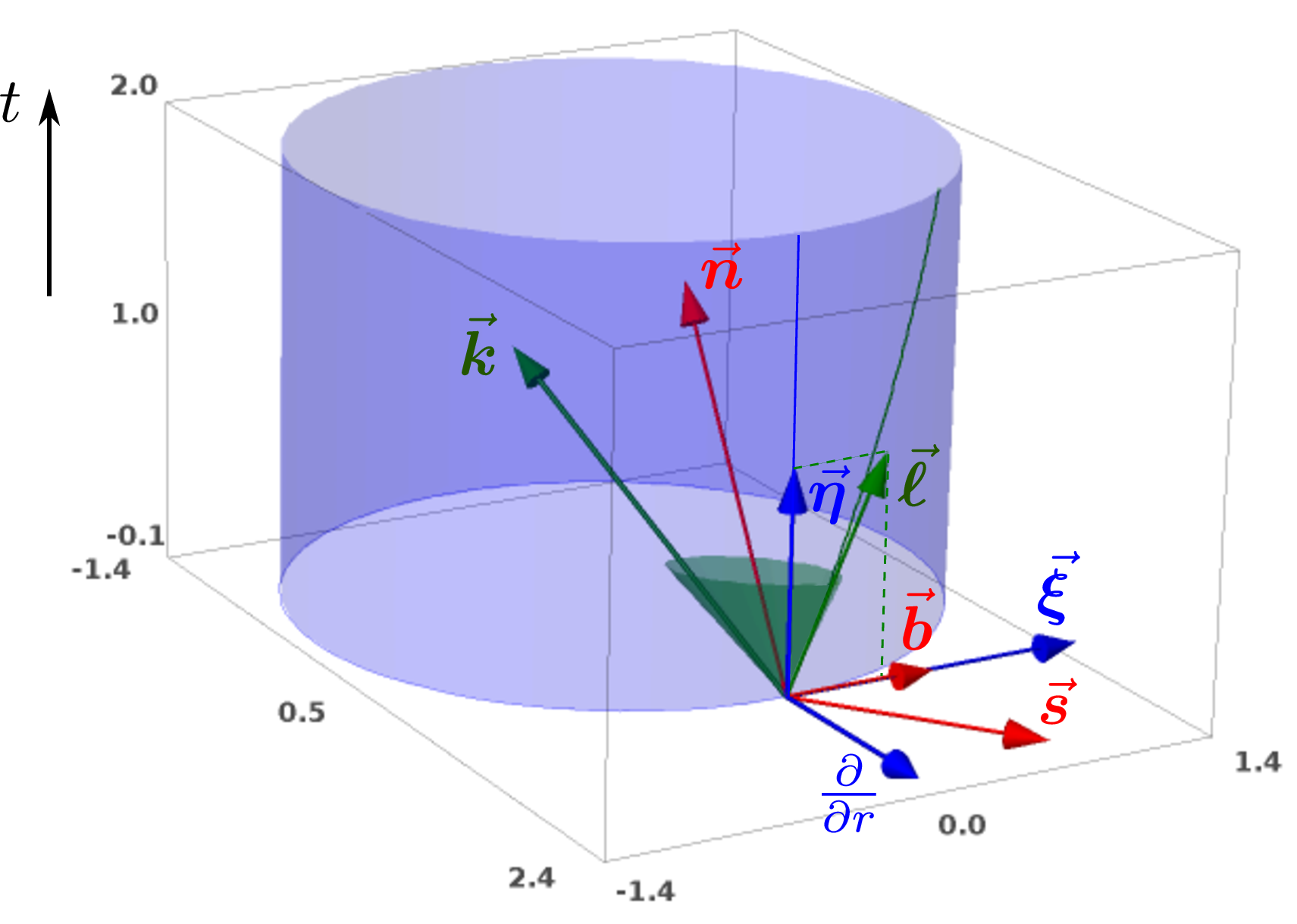}}
\caption{\label{f:horizon_vectors} \small
Spacetime diagram showing the event horizon of a Kerr black hole 
of angular momentum parameter $a/m=0.9$. This three-dimensional diagram is
cut at $\theta=\pi/2$ of the four-dimensional spacetime. 
The diagram is based on the 3+1 Kerr coordinates
$(t,r,\phi)$ described in Appendix~\ref{ap:Kerr} and the axes are labelled in units of $m$. 
The event horizon $\Hor$ is the blue cylinder of radius
$r= r_H = 1.435 m$ (this value results from $a=0.9m$ via (\ref{e:KE:r_H_def})) and the green cone is the future light cone at the 
point $(t=0, \theta=\pi/2, \phi=0)$ on $\Hor$.
The null vectors $\vw{\ell}$ and $\vw{k}$ (drawn in green) are tangent to this light cone, but not $\vw{\eta}$ which, although tangent to $\Hor$, being spacelike lies outside of the light cone. Note that relation (\ref{e:ell_eta_xi}) holds with 
$\omega_H = 0.313 m^{-1}$ (cf. Appendix~\ref{ap:Kerr}). The green line, to which 
$\vw{\ell}$ is tangent, is a null geodesic tangent to $\Hor$; if the figure was extended upward, it would show up as a helix.
$\vw{n}$ is the (timelike) unit normal to the hypersurface $t=0$. 
$\vw{s}$ is the (spacelike) unit normal to the 2-sphere $\Sps_0$ defined by
$t=0$ and $r=r_H$. Note that this 2-sphere is drawn here as a circle (the basis of the cylinder) because the dimension along $\theta$ has been suppressed.
The vector $\vw{b}$ is the unit vector along $\vw{\xi} = \partial/\partial\phi$. 
The vectors $(\vw{n},\vw{s},\vw{b})$ form an orthonormal basis (drawn in red) for the metric $\w{g}$. 
}
\end{figure}

The event horizon $\mathcal{H}$ is a null hypersurface; if it is stationary and 
axisymmetric, the symmetry generators $\vw{\eta}$ and $\vw{\xi}$ have to 
be tangent to it (cf. Fig.~\ref{f:horizon_vectors}). Moreover, 
any null normal $\vw{\ell}$ to $\Hor$ has to be a linear
combination of $\vw{\eta}$ and $\vw{\xi}$: up to some rescaling by a
constant factor, we may write
\be
\label{e:ell_eta_xi}
    \vw{\ell} = \vw{\eta} + \omega_{H} \vw{\xi} ,
\ee
where $\omega_{H}\geq 0$ is constant over $\Hor$ (rigidity theorem, cf. \cite{Carte79}) 
and is called the \emph{black-hole angular velocity}. 
Since $\omega_{H}$ is constant, $\vw{\ell}$ is itself a Killing 
vector and $\Hor$ is called a \emph{Killing horizon}. For a Kerr black hole of mass $m$
and angular momentum $a m$, we have $\omega_H = a/[2m r_H]$, where
$r_H=m+\sqrt{m^2-a^2}$ is the radius of the black-hole horizon. 
Since $\Hor$ is a null hypersurface, the normal $\vw{\ell}$ is null,
$\vw\ell\cdot\vw\ell = 0$.  For this reason, $\vw\ell$ is both normal
and tangent to $\Hor$. The field lines of $\vw{\ell}$ are null geodesics
tangent to $\Hor$; they are called the \emph{null generators} of
$\Hor$. One of them is drawn in Fig.~\ref{f:horizon_vectors}. 

Let $(x^\alpha)=(t,x^1,x^2,x^3)$ be a coordinate system on $\M$ that is
adapted to the stationarity, in the sense of (\ref{e:dodt_eta}), and 
regular on $\Hor$. In the case of a Kerr black hole, this means that $(x^\alpha)$ 
are not the standard Boyer-Lindquist coordinates, which are well known to 
be singular on $\Hor$. Regular coordinates on $\Hor$ are the Kerr coordinates,
either in their original version \citep{Kerr-1963} or in the 3+1 one,
and the Kerr-Schild coordinates, which are used in the numerical computations by 
\citet{Tchekhovskoy-2010,Tchekhovskoy-2011,McKinney-2012} discussed in 
Sec.~\ref{sect:mad}.  See Appendix~\ref{ap:Kerr} for more details on the
coordinate system and the coordinate representation of $\vw \ell$.

Then from (\ref{e:dodt_eta}) and (\ref{e:ell_eta_xi}),
$t$ is the parameter along the null geodesics generating $\Hor$ for which $\vw{\ell}$ 
is the tangent vector: 
\be
\label{e:ell_dxdt}
    \ell^\alpha = \frac{\D x^\alpha}{\D t} . 
\ee
(Note that in general $t$ is not an affine parameter along these geodesics.)
Since the coordinates $(t,x^i)$ are assumed regular on $\Hor$,
the 2-surfaces $\Sps_t$ of constant $t$ on $\Hor$ provide a regular slicing of $\Hor$ by a family 
of spacelike 2-spheres. Let us denote by $\vw{k}$ the future-directed null vector field defined on $\Hor$
by the following requirements (cf. Fig.~\ref{f:horizon_vectors}): 
\begin{enumerate}
\item $\vw{k}$ is orthogonal to $\Sps_t$,
\item  $\vw{k}$ obeys 
\be
\label{e:k_ell}
\vw{k}\cdot\vw{\ell} = -1.
\ee 
\end{enumerate}
Then, at each point of $\Sps_t$, $\mathrm{Span}(\vw{k},\vw{\ell})$ is the timelike 2-plane orthogonal
to $\Sps_t$. 
Note that $\vw{k}$ is transverse to $\Hor$ (i.e. is not tangent to it) and that, contrary to $\vw{\ell}$, 
the vector $\vw{k}$ depends on the choice of the coordinates $(t,x^i)$ (more precisely on the slicing $(\Sps_t)_{t\in\mathbb{R}}$ of $\Hor$, see e.g. \cite{Gourgoulhon-2005}).

The 2-surfaces $\Sps_t$ of constant $t$ on $\Hor$ are spacelike 2-spheres corresponding to what is commonly understood as
the ``black-hole surface", in analogy to ``stellar surface".

\subsection{Energy condition}
\label{sub-section-Energy-conditions}

Let $\w{T}$ be the energy-momentum tensor of matter and non-gravitational fields 
surrounding the black hole. We shall assume that it fulfills the 
so-called \emph{null energy condition} at the event horizon:
\be \label{e:Tll}
    \left. T_{\mu\nu} \ell^\mu \ell^\nu \right\vert _{\Hor} \geq 0  .
\ee
This is a very mild condition, which is satisfied by any ordinary matter
and any electromagnetic field. 
In particular, it follows (by some continuity argument timelike $\rightarrow$ null)
from the standard \emph{weak energy condition}
\citep{HawkiE73}, according to which energy measured locally by observers is always non-negative.

\section{Energy and angular-momentum conservation laws}
\label{s:conservation_laws}

In the mechanical Penrose process particles move on geodesics along which 
the energy $E$ and the angular
momentum $J$, as defined in Sec.~\ref{Section-Penrose-particles},
 are conserved quantities. Therefore they can be evaluated anywhere along the particle trajectories.
In particular at the black-hole surface where an energy flux can be calculated. In the general case of matter
with nongravitational interactions (e.g. a perfect fluid) or a field (e.g., electromagnetic) the energy
and angular momentum must be evaluated using the conservation equations and in such a case the fluxes of
the conserved quantities play the role equivalent to that of energy and angular momentum in the case of
particles.\footnote{In \citet{Abramowicz-2010} where generalizing the Penrose process was attempted, Eqs. (B.3) and
(B.4) are not correct because the ``energy at infinity" and ``angular momentum at infinity" that are used there, are not conserved
quantities}

\subsection{Energy conservation}
\label{s:energy_conservation}

Let us consider the ``energy-momentum density'' vector $\vw{P}$ defined by
\be
\label{e:def_P}
    P^\alpha = - T^\alpha_{\ \mu} \eta^\mu .
\ee
If matter and nongravitational fields obey the standard
\emph{dominant energy condition}\citep{HawkiE73} then $\vw{P}$ must be a future-directed
timelike or null vector as long as  $\vw{\eta}$ is timelike, i.e. outside
the ergoregion. In the ergoregion, where $\vw{\eta}$ is spacelike, 
there is no guarantee that $\vw{P}$ is timelike or null and even
 when it is timelike, $\vw{P}$ can be past-directed (an example in provided 
 in Fig.~\ref{f:negative_energy} below). 
Therefore, $\vw{P}$ cannot be interpreted as a physical energy-momentum density, hence
the quotes in the above denomination. 
Moreover, even outside the ergoregion, 
$\vw{P}$ does not correspond to the energy-momentum density measured by
any physical observer, since $\vw{\eta}$ fails to be some observer's 4-velocity, not being
a unit vector, except 
at infinity (cf. the discussion in Sec.~\ref{Section-Penrose-particles}). 
The vector $\vw{P}$ is known as the \emph{Noether current} associated 
with the symmetry generator $\vw{\eta}$ \citep{Szaba09, JaramG11}. It is conserved 
in the sense that 
\be
\label{e:P_conserved}
    \nabla_\mu P^\mu = 0 .
\ee
This is easily proved from the definition (\ref{e:def_P}) 
by means of (i) the energy-momentum conservation law 
$\nabla_\mu T^{\mu\nu} = 0$, (ii) the Killing equation obeyed by 
$\vw{\eta}$ and (iii) the symmetry of the tensor $\w{T}$. 
By Stokes' theorem, it follows from (\ref{e:P_conserved}) that
the flux of $\vw{P}$ through any closed\footnote{i.e. compact without boundary.} oriented hypersurface $\VV$ vanishes:
\be \label{e:flux_P_V}
    \oint_{\VV} \weps(\vw{P}) = 0,
\ee
where $\weps(\vw{P})$ stands for the 3-form obtained by setting $\vw{P}$ as the first argument of the Levi-Civita tensor $\weps$ (or volume 4-form) associated with the spacetime metric $\w{g}$:
\be \label{e:def_epsP}
    \weps(\vw{P}) := \weps(\vw{P},.,.,.) . 
\ee
In terms of components in a right-handed basis, 
\be \label{e:epsP_comp}
    \epsilon(\vw{P})_{\alpha\beta\gamma} = P^\mu \epsilon_{\mu\alpha\beta\gamma} =
\sqrt{-g} P^\mu [\mu,\alpha,\beta,\gamma], 
\ee
where $g := \det(g_{\alpha\beta})$ and $[\mu,\alpha,\beta,\gamma]$ is the alternating symbol
of four indices, i.e. $[\mu,\alpha,\beta,\gamma]=1$ ($-1$) if
$(\mu,\alpha,\beta,\gamma)$
is an even (odd) permutation of $(0,1,2,3)$, and $[\mu,\alpha,\beta,\gamma]=0$ 
otherwise.
Note that the integral (\ref{e:flux_P_V}) is intrinsically 
well defined, as the integral
of a 3-form over a three-dimensional oriented manifold. The proof of (\ref{e:flux_P_V})
relies on Stokes' theorem according to which the integral over $\VV$ is equal to 
the integral over the interior of $\VV$ of 
the exterior derivative of the 3-form $\weps(\vw{P})$; the latter being 
$\dd [ \weps(\vw{P}) ] = (\nabla_\mu P^\mu) \weps$, it 
vanishes identically as a consequence of (\ref{e:P_conserved}). 

\begin{figure}
\centerline{\includegraphics[height=0.20\textheight]{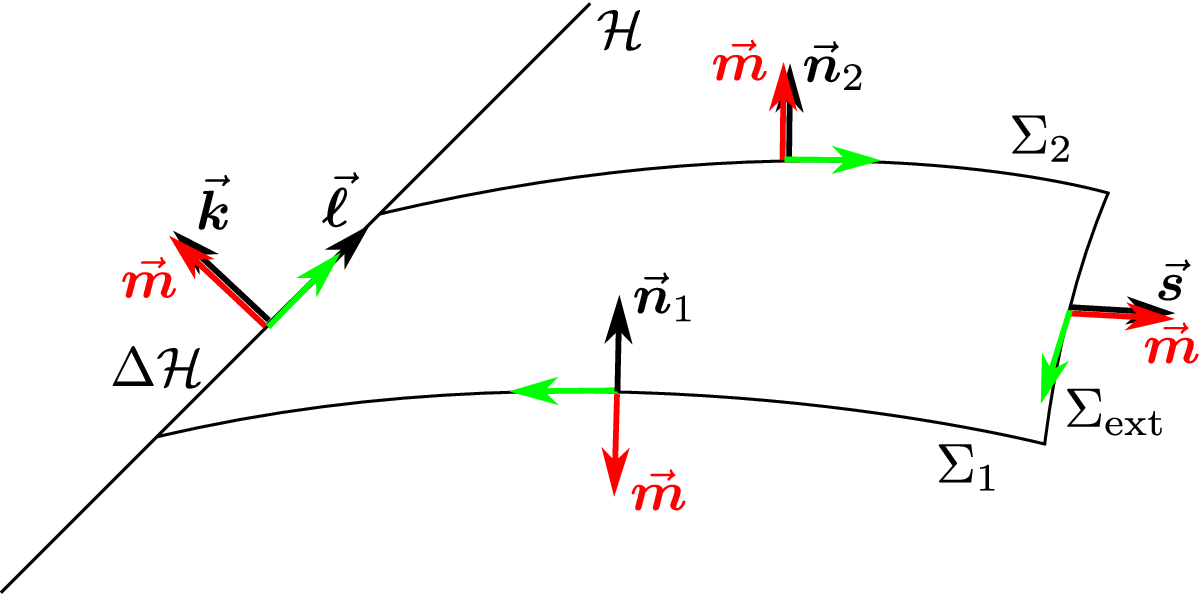}}
\caption{\label{f:hypersurfaces} \small
Closed hypersurface $\VV=\Sigma_1 \cup \Delta\Hor \cup \Sigma_2 \cup \Sigma_{\rm ext}$.
The green arrows depict the orientation of $\VV$, which is given by $\weps(\vw{m})$.}
\end{figure}

Let us apply (\ref{e:flux_P_V}) to the hypersurface $\VV$ 
defined as the following union:
\be \label{e:VV_union}
    \VV := \Sigma_1 \cup \Delta\Hor \cup \Sigma_2 \cup \Sigma_{\rm ext},
\ee
where (cf. Fig.~\ref{f:hypersurfaces})
\begin{itemize}
\item $\Sigma_1$ ($\Sigma_2$) is a compact spacelike hypersurface delimited by two 2-spheres,
$\Sp_1$ and $\Sp_1^{\rm ext}$ ($\Sp_2$ and $\Sp_2^{\rm ext}$), such that $\Sp_1$ ($\Sp_2$) lies on $\Hor$ and $\Sp_1^{\rm ext}$ (resp. $\Sp_2^{\rm ext}$) is located far from the black hole;
\item $\Sigma_2$ is assumed to lie entirely in the future of $\Sigma_1$; 
\item $\Delta\Hor$ is the portion of the event horizon $\Hor$ delimited by $\Sp_1$ and $\Sp_2$; 
\item $\Sigma_{\rm ext}$ is a timelike hypersurface having $\Sp_1^{\rm ext}$ and $\Sp_2^{\rm ext}$ for 
boundaries. 
\end{itemize}
We may choose, but this is not mandatory, the 2-spheres $\Sp_1$ and $\Sp_2$ to coincide with 
some slices of the foliation $(\Sps_t)_{t\in\mathbb{R}}$ of $\Hor$ mentioned in Sec.~\ref{sub-section-Kerr-horizon}:
$\Sp_1 = \Sps_{t_1}$ and $\Sp_2 = \Sps_{t_2}$. 

We choose the orientation of $\VV$ to be towards its exterior, 
but the final results do not depend upon this choice. The orientation of $\VV$
is depicted by the vector 
$\vw{m}$ in Fig.~\ref{f:hypersurfaces}. Note that this vector does not have to be normal to the various
parts of $\VV$ (in particular it is not normal to $\Delta\Hor$).  Its role is only to indicate that the orientation of $\VV$ is given by the 3-form $\weps(\vw{m})$ 
restricted to vectors tangent to $\VV$. More precisely, $\vw{m}$ is defined as follows: 
\begin{itemize}
\item on $\Sigma_1$, $\vw{m} = - \vw{n}_1$, the vector $\vw{n}_1$ being
the future-directed unit timelike normal to $\Sigma_1$;  
\item on $\Sigma_2$, $\vw{m} = \vw{n}_2$, the future-directed unit timelike normal to $\Sigma_2$; 
\item on $\Sigma_{\rm ext}$, $\vw{m} = \vw{s}$, the unit spacelike normal 
to $\Sigma_{\rm ext}$ oriented towards the exterior of $\VV$; 
\item on $\Delta\Hor$, $\vw{m}=\vw{k}$,  the future-directed null vector 
introduced above [cf. (\ref{e:k_ell})].
\end{itemize}
In view of (\ref{e:VV_union}), the property (\ref{e:flux_P_V}) gives
\be \label{e:flux_P_sum_int}
  \int_{\Sigma_1\downarrow} \weps(\vw{P}) \ 
  + \int_{{\ \atop\stackrel{\scriptstyle \Delta\Hor}{\scriptstyle \leftarrow}}} \weps(\vw{P}) \ 
  + \int_{\Sigma_2\uparrow} \weps(\vw{P}) \  
  + \int_{{\ \atop\stackrel{\scriptstyle \Sigma_{\rm ext}}{\scriptstyle \rightarrow}}} \weps(\vw{P})= 0 , 
\ee
where the arrows indicate the orientation (cf. Fig.~\ref{f:hypersurfaces}).
Let us then define the \emph{energy contained in $\Sigma_1$} by 
\bea \label{e:def_E1}
    E_1 &:=& \int_{\Sigma_1\uparrow} \weps(\vw{P}) 
        = - \int_{\Sigma_1} P_\mu n_1^\mu \, \D V \nonumber \\
        &=&  \int_{\Sigma_1} T_{\mu\nu} \eta^\mu n_1^\nu  \, \sqrt{\gamma}
        \, \D x^1 \, \D x^2 \, \D x^3  ,
\eea
the \emph{energy contained in $\Sigma_2$} by 
\bea\label{e:def_E2}
    E_2 &:=& \int_{\Sigma_2\uparrow} \weps(\vw{P})
        = - \int_{\Sigma_2} P_\mu n_2^\mu \, \D V  \nonumber \\
                &=& \int_{\Sigma_2} T_{\mu\nu} \eta^\mu n_2^\nu  \, \sqrt{\gamma}
        \, \D x^1 \, \D x^2 \, \D x^3  , 
\eea
the \emph{energy captured by the black hole between $\Sigma_1$ and $\Sigma_2$} by
\bea \label{e:E_H}
    \Delta E_H &:=& 
    \int_{{\ \atop\stackrel{\scriptstyle \Delta\Hor}{\scriptstyle \leftarrow}}} \weps(\vw{P})
    = - \int_{\Delta\Hor} P_\mu \ell^\mu \, \D V\nonumber \\
           & =& \int_{\Delta\Hor} T_{\mu\nu} \eta^\mu  \ell^\nu \, \sqrt{q} \, \D t \, \D y^1 \, \D y^2 
\eea
and the \emph{energy evacuated from the system between $\Sigma_1$ and $\Sigma_2$} 
by 
\bea\label{e:E_ext}
    \Delta E_{\rm ext} &:=& 
    \int_{{\ \atop\stackrel{\scriptstyle \Sigma_{\rm ext}}{\scriptstyle \rightarrow}}} \weps(\vw{P})
   =  \int_{\Sigma_{\rm ext}} P_\mu s^\mu \, \D V \nonumber \\
      & =& -  \int_{\Sigma_{\rm ext}} T_{\mu\nu} \eta^\mu s^\nu \, \sqrt{-h}
        \, \D t \, \D y^1 \, \D y^2  . 
\eea 
In the above formulas, 
\begin{itemize}
\item $\D V$ is the volume element induced on each hypersurface
by the spacetime Levi-Civita tensor $\weps$;
\item $(x^1,x^2,x^3)$ are generic coordinates
on $\Sigma_1$ and $\Sigma_2$ that are right-handed with respect
to the hypersurface orientation; 
\item $\gamma$ is the determinant of the  
components with respect to the coordinates $(x^1,x^2,x^3)$ of the 3-metric 
$\w{\gamma}$ induced by $\w{g}$ on $\Sigma_1$ or $\Sigma_2$;
\item 
$(t,y^1,y^2)$ are generic right handed coordinates on $\Sigma_{\rm ext}$; 
\item 
$h$ is the determinant of the  
components with respect to the coordinates $(t,y^1,y^2)$ of the 3-metric 
$\w{h}$ induced by $\w{g}$ on $\Sigma_{\rm ext}$ ($h<0$ since $\Sigma_{\rm ext}$
is timelike); 
\item $(t,y^1,y^2)$ are 
right-handed coordinates on $\Delta\Hor$ such that $t$ is the parameter 
along the null geodesics generating $\Hor$ associated with the null normal 
$\vw{\ell}$ (cf. (\ref{e:ell_dxdt}));
\item $q$ is the  determinant 
with respect to the coordinates $(y^1,y^2)$ of the 2-metric 
induced by $\w{g}$ on the 2-surfaces $t=\mathrm{const}$ in $\Delta\Hor$. 
\end{itemize}
The second and third equalities in
each of equations (\ref{e:def_E1})-(\ref{e:E_ext}) are established in Appendix~\ref{ap:flux_integrals}. 

With the above definitions, 
(\ref{e:flux_P_sum_int}) can be written as the energy conservation law
\be \label{e:ener_cons}
    {E_2 + \Delta E_{\rm ext} - E_1 = - \Delta E_H  } . 
\ee
Notice that the minus sign in front of $E_1$ arises from the change of orientation 
of $\Sigma_1$ between (\ref{e:flux_P_sum_int}) and the definition 
(\ref{e:def_E1}) of $E_1$.

\subsection{Angular-momentum conservation}
\label{sub:angmoment}
In a way similar to (\ref{e:def_P}), we define the angular-momentum density vector by 
\be \label{e:def_M}
    M^\alpha = T^\alpha_{\ \mu} \xi^\mu . 
\ee
Since $\vw{\xi}$ is a Killing vector, $\vw{M}$ obeys the conservation law
\be \label{e:divM}
    \nabla_\mu M^\mu = 0 . 
\ee
Let us introduce the 
\emph{angular momentum contained in $\Sigma_1$} and that 
\emph{contained in} $\Sigma_2$ by 
\bea
\label{e:def_J1}
    J_1& :=& \int_{\Sigma_1\uparrow} \weps(\vw{M}) = - \int_{\Sigma_1} M_\mu n_1^\mu \, \D V \nonumber \\
       & =& - \int_{\Sigma_1} T_{\mu\nu} \xi^\mu n_1^\nu  \, \sqrt{\gamma}
        \, \D x^1 \, \D x^2 \, \D x^3  
\eea
and
\bea
    J_2 &:=& \int_{\Sigma_2\uparrow} \weps(\vw{M}) = - \int_{\Sigma_2} M_\mu n_2^\mu \, \D V \nonumber \\
        &= &- \int_{\Sigma_2} T_{\mu\nu} \xi^\mu n_2^\nu  \, \sqrt{\gamma}
        \, \D x^1 \, \D x^2 \, \D x^3,
\eea
the \emph{angular momentum captured by the black hole between $\Sigma_1$ and $\Sigma_2$} by
\bea \label{e:J_H}
    \Delta J_H &:=& 
    \int_{{\ \atop\stackrel{\scriptstyle \Delta\Hor}{\scriptstyle \leftarrow}}} \weps(\vw{M})
        = - \int_{\Delta\Hor} M_\mu \ell^\mu \, \D V\nonumber \\
        &=& - \int_{\Delta\Hor} T_{\mu\nu} \xi^\mu  \ell^\nu \, \sqrt{q} \, \D t \, \D y^1 \, \D y^2
\eea
and the \emph{angular momentum evacuated from the system between $\Sigma_1$ and $\Sigma_2$} by
\bea \label{e:def_Jext}
    J_{\rm ext} &:=& 
    \int_{{\ \atop\stackrel{\scriptstyle \Sigma_{\rm ext}}{\scriptstyle \rightarrow}}}
    \weps(\vw{M})
    = \int_{\Sigma_{\rm ext}} M_\mu s^\mu \, \D V \nonumber \\
       & =& \int_{\Sigma_{\rm ext}} T_{\mu\nu} \xi^\mu s^\nu \, \sqrt{-h}
        \, \D t \, \D y^1 \, \D y^2 . 
\eea 
We deduce then from (\ref{e:divM}) that, similarly to (\ref{e:ener_cons}), 
\be \label{e:angu_mom}
   {J_2 + J_{\rm ext} - J_1 = - \Delta J_H  } . 
\ee

\subsection{Explicit expressions in adapted coordinates}
\label{s:adapted_coord}

Let us call \emph{adapted coordinates} any right-handed spherical-type coordinate system
$(x^\alpha) = (t,r,\th,\phi)$ such that (i)
$t$ and $\phi$ are associated with the two spacetime symmetries, so that the two independent Killing vectors
are $\vw{\eta} = \partial/\partial t$ and $\vw{\xi} = \partial/\partial \phi$,
(ii) the event horizon $\Hor$ is the hypersurface defined by $r=\mathrm{const} = r_H$, 
(iii) the timelike hypersurface $\Sigma_{\rm ext}$ is defined by
$r=\mathrm{const}=r_{\rm ext}$ and $t\in[t_1,t_2]$, where $t_1$ and $t_2$
are two constants such that $t_1<t_2$ and (iv) the spacelike hypersurface $\Sigma_1$
($\Sigma_2$) is defined by $t=t_1$ and $r\in[r_H, r_{\rm ext}]$
($t=t_2$ and $r\in[r_H, r_{\rm ext}]$). 
Then $\Delta\Hor$ is the hypersurface defined by $r=r_H$ and 
$t\in[t_1, t_2]$. 
In the case of Kerr spacetime, an example of adapted coordinates are the
3+1 Kerr coordinates described in Appendix~\ref{ap:Kerr}. 

On $\Sigma_1$ or $\Sigma_2$, $(r,\th,\phi)$ are coordinates that are right-handed 
with respect to the ``up'' orientation of these hypersurfaces used in the
definitions (\ref{e:def_E1})-(\ref{e:def_E2}) of $E_1$ and $E_2$. Consequently, 
\bea
        E_{1,2} &=& \int_{\Sigma_{1,2}} \epsilon(P)_{r\th\phi} \, \D r\, \D\th \, \D\phi\nonumber \\
            &=& \int_{\Sigma_{1,2}} \sqrt{-g} P^t \, \underbrace{[t, r, \th,\phi]}_{1}\nonumber 
                \, \D r\, \D\th \, \D\phi , 
\eea
where the second equality results from (\ref{e:epsP_comp}). 
Now, (\ref{e:def_P}) yields  $P^t = - T^t_{\ \, \mu}\eta^\mu = -T^t_{\ \, t}$
since $\eta^\alpha = (1,0,0,0)$ in adapted coordinates.
We conclude that 
\bea \label{e:E_1_E_2_adapted}
    E_1 &=&  - \int_{\Sigma_1} \, T^t_{\ \, t} \, \sqrt{-g}
        \, \D r \, \D \th \, \D \phi \nonumber \\
    \quad\mbox{and}\quad&& \nonumber \\
     E_2 &=& -  \int_{\Sigma_2} \, T^t_{\ \, t} \, \sqrt{-g}
        \, \D r \, \D \th \, \D \phi . 
\eea
As a check, we note that the above formulas can also be recovered from the 
expressions involving $T_{\mu\nu}\eta^\mu n_{1,2}^\nu$ in 
(\ref{e:def_E1})-(\ref{e:def_E2}). Indeed, the unit timelike normal $\vw{n}$ to
$\Sigma_1$ or $\Sigma_2$ obeys $n_\alpha = (-N,0,0,0)$, where 
$N$ is the lapse function of the spacetime foliation by $t=\mathrm{const}$
hypersurfaces (see e.g. \cite{Gourg12}). Accordingly 
$T_{\mu\nu} \eta^\mu n^\nu = T^\nu_{\ \, \mu} \eta^\mu n_\nu = T^t_{\ \, t} (-N)$. 
Since $N\sqrt{\gamma} = \sqrt{-g}$, we get (\ref{e:E_1_E_2_adapted}). 

On $\Delta\Hor$, $(t,\th,\phi)$ are coordinates that are right handed 
with respect to the ``inward'' orientation used in the definition 
(\ref{e:E_H}) of $\Delta E_H$. Indeed
\bea
\weps(\vw{m},\vw{\partial}_t, \vw{\partial}_\th, \vw{\partial}_\phi) &=& 
\weps(\vw{k},\vw{\partial}_t, \vw{\partial}_\th, \vw{\partial}_\phi)\nonumber \\
 &=& k^r \epsilon_{rt\th\phi} = - \underbrace{k^r}_{<0}  
 \underbrace{\epsilon_{tr\th\phi}}_{>0} > 0 .  \nonumber \\
\eea
Accordingly, 
\bea
    \Delta E_H &=& \int_{\Delta\Hor} \epsilon(P)_{t\th\phi} \, \D t\, \D\th \, \D\phi\nonumber \\
       & =& \int_{\Delta\Hor} \sqrt{-g} P^r \, \underbrace{[r, t, \th,\phi]}_{-1}
                \, \D t\, \D\th \, \D\phi , \nonumber \\
\eea
where the second equality results from (\ref{e:epsP_comp}). 
Since $P^r =  -T^r_{\ \, t}$ from (\ref{e:def_P}), we get 
\be \label{e:E_H_adapted}
    \Delta E_H = \int_{\Delta\Hor} T^r_{\ \, t} \, \sqrt{-g} \, \D t \, \D \theta 
    \, \D \phi .
\ee

On $\Sigma_{\rm ext}$, it is $(t,\phi,\th)$, and not $(t,\th,\phi)$, that 
constitutes a right-handed coordinate system with respect to the 
orientation used in the definition (\ref{e:E_ext}) of $\Delta E_{\rm ext}$. 
Indeed
\bea
\weps(\vw{m},\vw{\partial}_t, \vw{\partial}_\phi, \vw{\partial}_\th) &=& 
\weps(\vw{s},\vw{\partial}_t, \vw{\partial}_\phi, \vw{\partial}_\th)\nonumber \\
& =& s^r \epsilon_{rt\phi\theta} = \underbrace{s^r}_{>0}  
 \underbrace{\epsilon_{tr\th\phi}}_{>0} > 0 .  \nonumber \\
\eea
We have therefore
\bea
    \Delta E_{\rm ext} &=& \int_{\Sigma_{\rm ext}} \epsilon(P)_{t\phi\th} \, \D t\, \D\th \, \D\phi\nonumber \\
        &=& \int_{\Sigma_{\rm ext}}  \sqrt{-g} P^r \, \underbrace{[r, t, \phi,\th]}_{1}
                \, \D t\, \D\th \, \D\phi , \nonumber \\
\eea
Substituting $-T^r_{\ \, t}$ for $P^r$, we get 
\be \label{e:E_ext_adapted}
\Delta E_{\rm ext} 
    = -  \int_{\Sigma_{\rm ext}} T^r_{\ \, t} \, \sqrt{-g}
        \, \D t \, \D \th \, \D \phi . 
\ee

The formulas for the angular momentum are similar to the above ones, 
with $T^t_{\ \, t}$ replaced by $-T^t_{\ \, \phi}$
and $T^r_{\ \, t}$ replaced by $-T^r_{\ \, \phi}$:
\bea \label{e:J_1_J_2_adapted}
    J_1 &=&  \int_{\Sigma_1} \, T^t_{\ \, \phi} \, \sqrt{-g}
        \, \D r \, \D \th \, \D \phi \nonumber \\
    \quad\mbox{and}\quad &&\nonumber \\
     J_2 &=& \int_{\Sigma_2} \, T^t_{\ \, \phi} \, \sqrt{-g}
        \, \D r \, \D \th \, \D \phi , 
\eea
\be \label{e:J_H_adapted}
    \Delta J_H = - \int_{\Delta\Hor} T^r_{\ \, \phi} \, \sqrt{-g} \, \D t \, \D \theta 
    \, \D \phi ,
\ee
\be \label{e:J_ext_adapted}
\Delta J_{\rm ext} 
    = \int_{\Sigma_{\rm ext}} T^r_{\ \, \phi} \, \sqrt{-g}
        \, \D t \, \D \th \, \D \phi . 
\ee

Expressions (\ref{e:E_1_E_2_adapted})-(\ref{e:E_ext_adapted})
and (\ref{e:J_1_J_2_adapted})-(\ref{e:J_ext_adapted}), as well as the 
energy conservation law (\ref{e:ener_cons})  and the angular-momentum conservation 
law (\ref{e:angu_mom}), are rederived in Appendix~\ref{ap:spherical}, via a pure
coordinate-based calculation.

\section{General conditions for black-hole rotational energy extraction}
\label{section-General}

\subsection{General case}

For definiteness, let us consider that $\Sigma_1$ and $\Sigma_2$ are parts of a foliation of spacetime
by a family of spacelike hypersurfaces $(\Sigma_t)_{t\in\mathbb{R}}$:
\be
    \Sigma_1 = \Sigma_{t_1} \quad\mbox{and}\quad \Sigma_2 = \Sigma_{t_2} \quad\mbox{with}\quad t_2 > t_1 . 
\ee
For instance, in the case of a Kerr black hole, the hypersurface label $t$ can be chosen to be the Kerr-Schild time coordinate introduced in Appendix~\ref{ap:Kerr}. 

In (\ref{e:ener_cons}), we may then interpret $E_1$ as the ``initial energy'', i.e. the energy
``at the time $t_1$'', 
$E_2$ as the ``final energy'', i.e. the energy ``at the time $t_2$'' and
$\Delta E_{\rm ext}$ as the energy evacuated from the system between the times $t_1$ and $t_2$.
Accordingly, the ``energy gained by the world outside of the black hole'' between  $t_1$ and $t_2$ is 
defined as 
\be \label{e:energy_gain}
    \Delta E := E_2 + \Delta E_{\rm ext} - E_1 . 
\ee
Then, energy will be extracted from the black hole if, and only if  $\Delta E > 0$.
In view of the conservation law (\ref{e:ener_cons}), we conclude that energy is extracted from a black hole if, and only if, 
\be  \label{e:Penrose_process}
    { \Delta E_H < 0 } . 
\ee
We refer to any process that accomplishes this as a \emph{Penrose process}.

Let us assume that the energy-momentum tensor obeys the \emph{null energy condition} 
(cf. Sect. \ref{sub-section-Energy-conditions}) on the event horizon: 
$\left. T_{\mu\nu} \ell^\mu \ell^\nu \right| _{\Hor} \geq 0 $ [Eq.~(\ref{e:Tll})]. 
As mentioned above, this is a rather mild condition, implied by the standard weak energy condition.
From (\ref{e:ell_eta_xi}), (\ref{e:def_P}) and (\ref{e:def_M}), it follows that
\[
    T_{\mu\nu} \ell^\mu \ell^\nu = T_{\mu\nu}(\eta^\nu + \omega_{H} \xi^\nu) \ell^\mu
        = - P_\mu \ell^\mu + \omega_{H} \,  M_\mu \ell^\mu . 
\]
Integrating (\ref{e:Tll}) over $\Delta\Hor$ yields then 
\be
    - \int_{\Delta\Hor} P_\mu \ell^\mu \, \D V + \omega_{H} \int_{\Delta\Hor} M_\mu \ell^\mu \, \D V 
        \geq 0 , 
\ee
where we have used the fact that $\omega_{H}$ is constant. 
Using (\ref{e:E_H}) and (\ref{e:J_H}), the above relation can be rewritten as 
$\Delta E_H - \omega_{H} \Delta J_H \geq 0$,
i.e. 
\be \label{e:omegJ_EH}
    { \omega_{H} \Delta J_H \leq \Delta E_H } . 
\ee
In view of (\ref{e:omegJ_EH}) and $\omega_{H} \geq 0$, the black-hole energy extraction condition 
(\ref{e:Penrose_process}) implies
\be
\label{e:Penrose_am}
    {\Delta J_H < 0} . 
\ee
We conclude the following:
\begin{quote}
For a matter distribution
or a nongravitational field obeying the null energy condition, 
a necessary and sufficient condition for energy extraction from a rotating black 
hole is that it absorbs negative energy $\Delta E_H$ and negative angular momentum $\Delta J_H$. 
\end{quote}
\begin{figure}
\centerline{\includegraphics[width=0.45\textwidth]{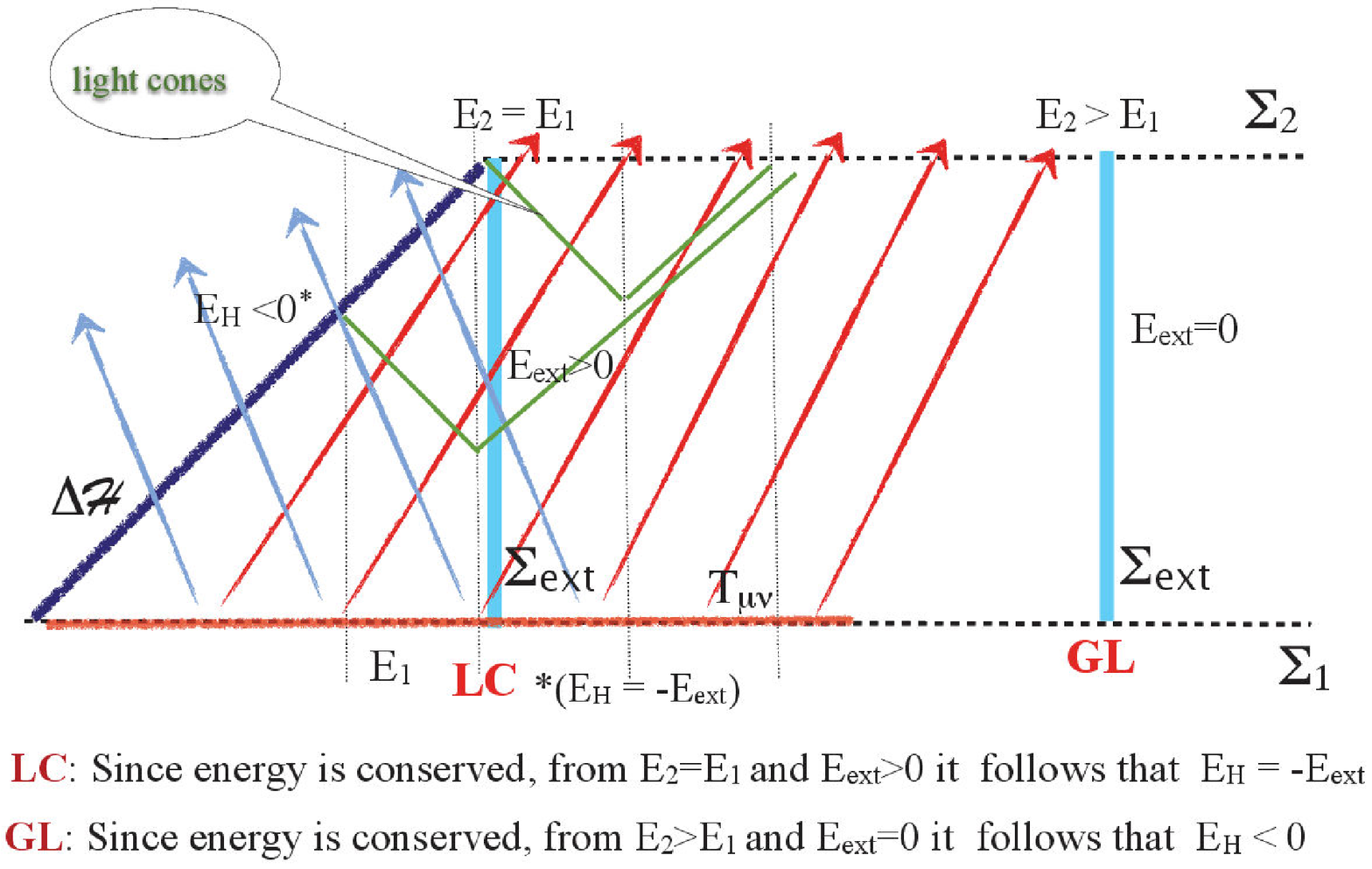}}
\centerline{\includegraphics[width=0.45\textwidth]{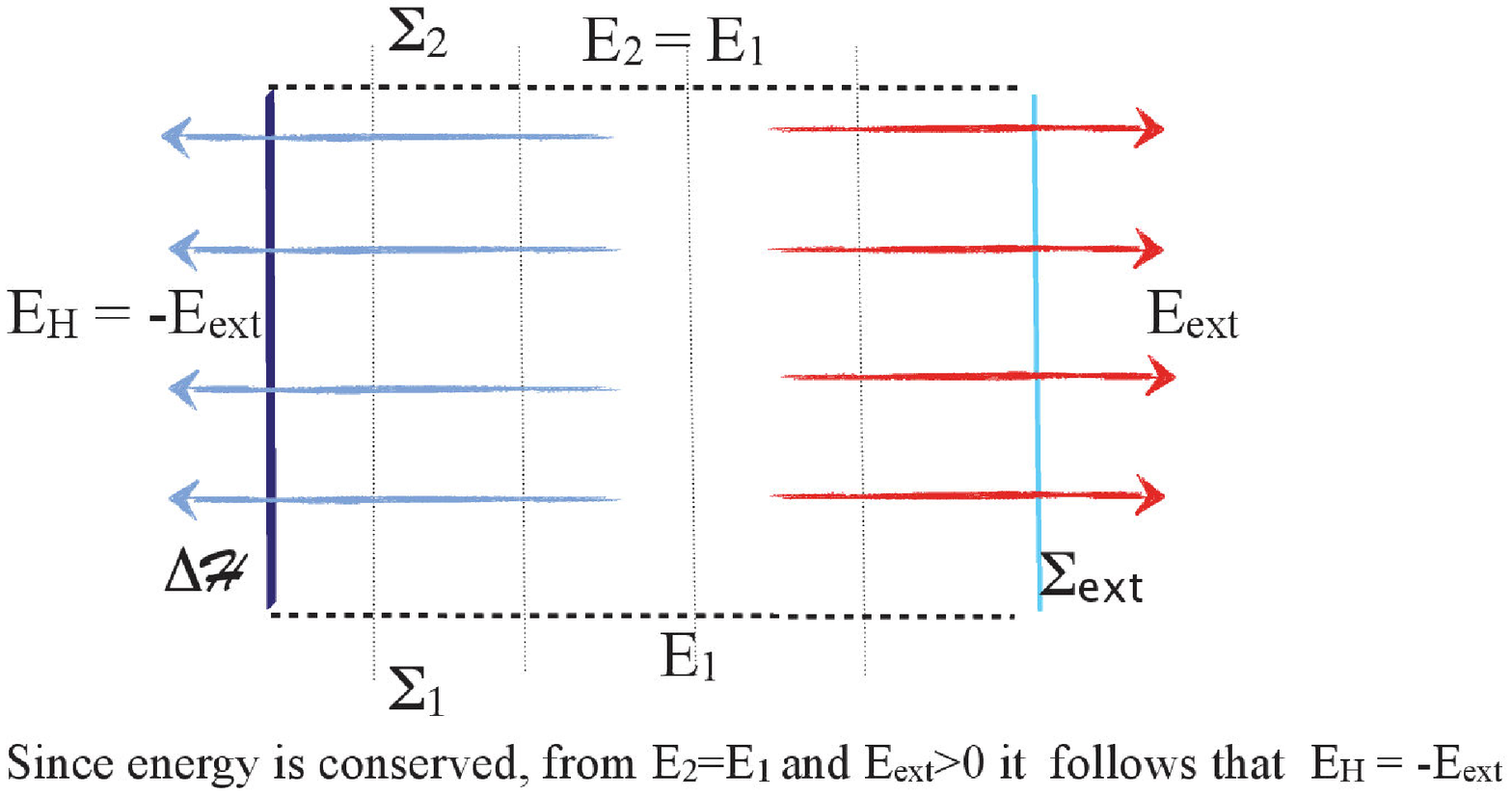}}
\caption{\label{f:Econs} \small Two views of the energy balance in a Penrose process. {\sl Top}: Global (GL)  with $E_2 > E_1$
and $\Delta E_{\rm ext}=0$.
{\sl Bottom}: local (LC) stationary view with $E_2 = E_1$ but $\Delta E_{\rm ext}= - \Delta E_H >0$. The region of spacetime concerned
with this view is marked ``LC" on the top figure.}
\end{figure}
Eqs. (\ref{e:Penrose_process}), (\ref{e:omegJ_EH}) and (\ref{e:Penrose_am}) are identical with
Eqs. (\ref{eq:Penr_E}), (\ref{eq:Penr_Epos}) and (\ref{eq:Penr_J}) describing the condition for the 
Penrose process. They describe the same physics: in order to extract energy from a rotating black
hole one must feed it negative energy and angular momentum.

{\sl Any extraction of black hole's rotational energy by interaction with matter and/or (nongravitational) fields is a Penrose
process.}

\subsection{Penrose process in terms of the Noether current $\vw{P}$}
\label{s:Penrose_Noether_current}

Given the expression (\ref{e:E_H}) of $\Delta E_H$, we note that the
Penrose-process condition 
(\ref{e:Penrose_process}) implies $P_\mu \ell^\mu > 0$ on some part 
of $\Delta\Hor$. Since $\vw{\ell}$ is a future-directed null vector, 
$P_\mu \ell^\mu > 0$ if, and only if, $\vw{P}$ is either (i) spacelike or 
(ii) past directed timelike or past directed null. Therefore, we conclude that 
\begin{quote}
A necessary condition for a Penrose process to occur is to have
the Noether current $\vw{P}$ be spacelike or 
past directed (timelike or null) on some part of $\Delta\Hor$. 
\end{quote}
As we already noticed in Sec.~\ref{s:energy_conservation}, if the matter 
or fields fulfil the standard dominant energy condition, the vector 
$\vw{P}$ is always future directed timelike or null outside the ergoregion; 
therefore it can be spacelike or past directed only in the ergoregion.

\subsection{Applications of the Penrose-process energy balance}
\label{subsect: applications}

The energy balance equations derived above can be applied to basically two views of energy extraction from a black hole.
First, one can use  global (GL) spacetime view applied
to theoretically described ``real" astrophysical systems (Fig. \ref{f:Econs} - top). Matter and/or fields have limited space extent, the timelike hypersurface $\Sigma_{\rm ext}$ is placed sufficiently far so that $\Delta E_{\rm ext}=0$. When there is energy extraction,
i.e. when $\Delta E >0$, then $E_2 > E_1$. This is the view we will have in mind in 
Secs.~\ref{s:examples} and \ref{sect:em}. 

When dealing with numerical simulations, however, such global view is usually unpractical. The simulation is performed in a box of limited size and
the system is brought to stationary state. The view presented in the bottom part of Fig. \ref{f:Econs} is then more adapted to the energy balance.
Because of stationarity one has $E_2 = E_1$ but $\Delta E_{\rm ext}>0$. When the numerical code conserves energy very well, the energy balance implies
$\Delta E_H<0$. This is the view applied in Sec. \ref{sect:mad}.

\section{Various examples of the Penrose process}
\label{s:examples}

In what follows we will apply Eqs. (\ref{e:def_E1}) to (\ref{e:ener_cons}) and  (\ref{e:def_J1}) to (\ref{e:angu_mom})
to various black-hole plus matter (or fields) configurations. We first show that 
in the case of particles one recovers the
standard Penrose-process formulae. Then we shall apply our formalism to the cases of
a scalar field and a perfect fluid. 
The case of the electromagnetic field is treated in Sec.~\ref{sect:em}. 

\subsection{Mechanical Penrose-process test}
\label{sub:mechP}

Let us show that the formalism developed above reproduces the mechanical Penrose 
process for a single particle that 
breaks up into two fragments in the ergoregion. 

The energy-momentum tensor of a massive particle of mass
$\mathfrak{m}$ and 4-velocity $\vw{u}$ is (cf. e.g. \cite{PoissPV11})
\bea \label{e:T_particle}
    T_{\alpha\beta}(M) = \mathfrak{m} && \int_{-\infty}^{+\infty} 
        \delta_{A(\tau)}(M)\; 
        g_\alpha^{\ \, \mu}(M,A(\tau)) u_\mu(\tau) \; \nonumber \\
     &&   \qquad \qquad \times 
        g_\beta^{\ \, \nu}(M,A(\tau))    u_\nu(\tau) \; \D\tau , 
\eea
where $M\in\M$ is the spacetime point at which $T_{\alpha\beta}$ is evaluated, 
$\tau$ stands for the particle's proper time, $A(\tau)\in\M$ is the spacetime point 
occupied by the particle at the proper time $\tau$,
$g_\alpha^{\ \, \mu}(M,A)$ is the parallel propagator from the point $A$ to the point $M$ along 
the unique geodesic\footnote{Thanks to the Dirac distribution in (\ref{e:T_particle}), 
only the limit $M\rightarrow A$ matters, so that we can assume that there is a unique geodesic 
connecting $A$ to $M$.} connecting $A$ to $M$ (cf. Sec.~5 of \cite{PoissPV11} or Appendix~I of \cite{Carro04}) 
and $\delta_{A}(M)$ is the Dirac distribution on $(\M,\w{g})$ centered at the point $A$: it is 
defined by the identity 
\be
    \int_{\mathscr{U}} \delta_A(M) f(M) \, \sqrt{-g} \, \D^4 x = f(A) , 
\ee
for any four-dimensional domain $\mathscr{U}$ around $A$ and any scalar field $f:\;\mathscr{U}\rightarrow\mathbb{R}$. 
In terms of a coordinate system $(x^\alpha)$ around
$A$:
\be \label{e:delta_A_M_explicit}
    \delta_A(M) = \frac{1}{\sqrt{-g}} \, \delta(x^0-z^0) \,\delta(x^1-z^1) \,\delta(x^2-z^2) \,\delta(x^3-z^3) ,  
\ee
where $\delta$ is the standard Dirac distribution on $\mathbb{R}$, $(x^\alpha)$ are the coordinates
of $M$, $(z^\alpha)$ those of $A$ and $g$ is the determinant of the components of 
the metric tensor with respect to the coordinates $(x^\alpha)$. 

\begin{figure}
\centerline{\includegraphics[height=0.20\textheight]{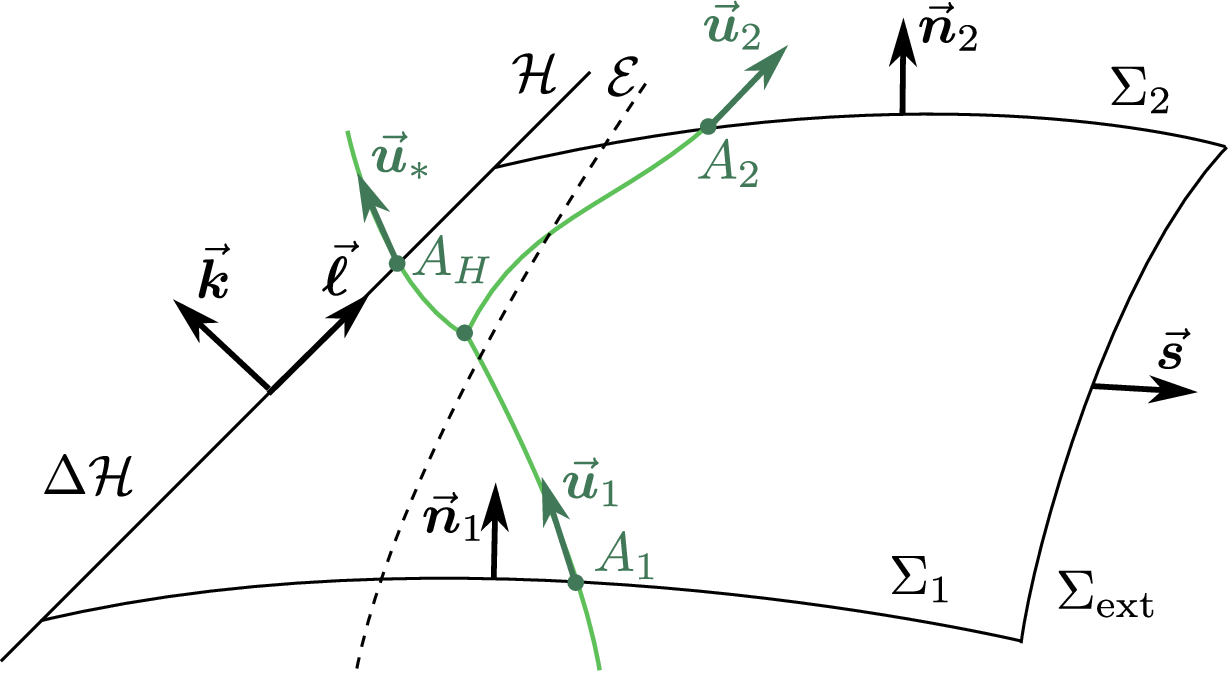}}
\caption{\label{f:particle} \small
Penrose process for a particle. The dashed line $\mathcal{E}$ marks the ergosphere.}
\end{figure}

The Noether current  corresponding to (\ref{e:T_particle}) is
formed via (\ref{e:def_P}):
\bea \label{e:P_particle}   
     P_\alpha(M) &=&  \mathfrak{m} \int_{-\infty}^{+\infty} \delta_{A(\tau)}(M) 
        \left[ -  \;
        g_\sigma^{\ \, \nu}(M,A(\tau))  u_\nu(\tau) 
     \eta^\sigma(M) \right] \nonumber \\
     && \quad  \quad \quad  \quad  \quad \quad \times g_\alpha^{\ \, \mu}(M,A(\tau)) u_\mu(\tau) \; \D\tau . 
\eea
This means that $\vw{P}$ is a distribution vector whose support
is the particle's worldline and that is collinear to the particle's 4-velocity. 

Let us choose $\Sigma_1$ and $\Sigma_2$ such that $\Sigma_1$ encounters the original particle $\Pp_1$
(mass $\mathfrak{m}_1$, 4-velocity $\vw{u}_1$) at the event $A_1$,
$\Sigma_2$ encounters the escaping fragment $\Pp_2$ (mass $\mathfrak{m}_2$, 4-velocity $\vw{u}_2$) at the
event $A_2$ and the infalling fragment $\Pp_*$ (mass $\mathfrak{m}_*$, 4-velocity $\vw{u}_*$) crosses
the horizon on $\Delta\Hor$, at the event $A_H$ (cf. Fig.~\ref{f:particle}). 
By plugging (\ref{e:T_particle}) into (\ref{e:def_E1}), we get
\begin{eqnarray}
    E_1 &= &\mathfrak{m}_1 \int_{\Sigma_1} \int_{-\infty}^{+\infty} 
        \delta_{A(\tau)}(M)\; g_\mu^{\ \, \rho}(M,A(\tau)) (u_1)_\rho(\tau) \;\nonumber \\
   && \qquad \qquad   \times  g_\nu^{\ \, \sigma}(M,A(\tau))  (u_1)_\sigma(\tau)  \;  
              \eta^\mu(M) \, n_1^\nu(M)  \nonumber \\
        &&   \qquad \qquad  \qquad \qquad \qquad \times   \sqrt{\gamma}
        \D x^1 \, \D x^2 \, \D x^3\, \D\tau .\label{e:Part_E1}
\end{eqnarray}

This formula (see Appendix~\ref{ap:Part_sp}) can be reduced to
\be \label{e:E_1_particle}
    E_1 = - \mathfrak{m}_1  \left. (\eta_\mu u_1^\mu) \right| _{A_1} = - \mathfrak{m}_1 \, \eta_\mu u_1^\mu , 
\ee
where the second equality stems from the fact that $\eta_\mu u_1^\mu$ is constant along 
$\Pp_1$'s worldline, since the latter is a geodesic and $\vw{\eta}$ is a Killing vector. 
That $\Pp_1$'s worldline is a geodesic follows from the energy-momentum conservation 
law $\nabla_\mu T^{\alpha\mu}=0$ with the form (\ref{e:T_particle}) for the energy-momentum tensor
(see Sec.~19.1 of \cite{PoissPV11} for details). 
We recover in (\ref{e:E_1_particle}) the standard expression of the energy involved in textbook 
discussions of the Penrose process (see
\cite{Carro04,Hartle-2003,Wald-1984} and Sec.~\ref{Section-Penrose-particles}). 

Similarly, for the outgoing particle one gets
\be
    E_2 = - \mathfrak{m}_2\, \eta_\mu u_2^\mu . 
\ee

For the particle crossing the horizon, by
plugging (\ref{e:T_particle}) with the characteristics of the infalling fragment $\Pp_*$ into (\ref{e:E_H}), we get
\bea 
 \Delta E_H &=& \mathfrak{m}_* \int_{\Delta\Hor} \int_{-\infty}^\infty \delta_{A(\tau)}(M) \,  (u_*)_\mu(\tau) \eta^\mu(M) \;(u_*)_\nu(\tau)\nonumber \\
 && 
\qquad \qquad\qquad \quad \times  \ell^\nu(M) \, \sqrt{q} \, \D t \, \D y^1 \, \D y^2 \, \D\tau . \label{e:Part_EH}
\eea
As shown in Appendix~\ref{ap:Part_null} this can be reduced to
\be \label{e:E_H_particle}
    \Delta E_H = - \mathfrak{m}_* \left. (\eta_\mu u_*^\mu) \right| _{A_H} = - \mathfrak{m}_* \, \eta_\mu u_*^\mu . 
\ee
As for $\Pp_1$ and $\Pp_2$, the independence of $\eta_\mu u_*^\mu$ from the specific point of 
$\Pp_*$'s worldline where it is evaluated results from the fact that $\Pp_*$'s worldline is
a geodesic.

\begin{figure}
\centerline{\includegraphics[height=0.3\textheight]{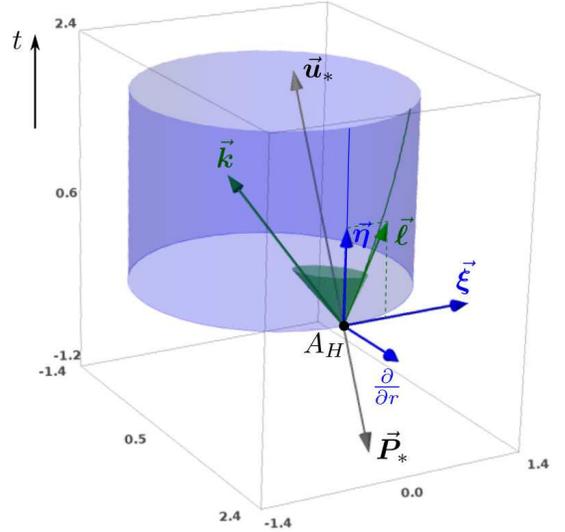}}
\caption{\label{f:negative_energy} \small
Spacetime diagram showing the 4-velocity $\vw{u}_*$ and the energy-momentum
density vector $\vw{P}_*$ of a negative-energy particle $\Pp_*$ entering 
the event horizon of a Kerr black hole 
of angular-momentum parameter $a/m=0.9$
(see Figs.~\ref{f:horizon_vectors} and \ref{f:particle}). At the horizon, the particle is
characterized by the following coordinate velocity: $\D r/\D t = -0.32$, 
$\D\th/\D t = 0$, and $\D\phi/\D t = -0.18 \omega_H$, resulting in the
4-velocity $u_*^\alpha = (2.38, -0.76,0,-0.13)$ and in the positive scalar 
product $\eta_\mu u_*^\mu = 0.042$. The ``vector'' $\vw{P}_*$, which is actually
a distribution, is drawn with an arbitrary scale. 
}
\end{figure}

Finally, in the present case, we have clearly $\Delta E_{\rm ext} = 0$. 
Therefore the energy gain formula (\ref{e:energy_gain}) reduces to 
$\Delta E = E_2 - E_1$ and we recover the standard Penrose process
discussed in Sec.~\ref{Section-Penrose-particles}: $E_2 > E_1$ if, and only if, 
$\Delta E_H < 0$, i.e., if and only if $\eta_\mu u_*^\mu > 0$. This is possible only in the ergoregion, where the Killing vector
$\vw{\eta}$ is spacelike.  
Note that  $\eta_\mu u_*^\mu > 0$ implies that the term in square brackets in 
(\ref{e:P_particle}) is negative, so that the Noether current
$\vw{P}_*$ of $\Pp_*$ is a timelike vector (being collinear to $\vw{u}_*$)
that is \emph{past directed}. This is in agreement with 
the statement made in Sec.~\ref{s:Penrose_Noether_current} and 
is illustrated in Fig.~\ref{f:negative_energy}.

\subsection{Scalar field (super-radiance)} 
\label{s:scalar_field}

Let us consider a complex scalar field $\Phi$ ruled by the standard
Lagrangian
\be
    \mathcal{L} = - \frac{1}{2} \left[
    \nabla_\mu \bar\Phi \nabla^\mu \Phi + V(|\Phi|^2) \right] ,
\ee
where  $\bar\Phi$ stands for $\Phi$'s complex conjugate and
$V(|\Phi|^2)$ is some potential ($V(|\Phi|^2) = \left({\mathfrak{m}}/{\hbar}\right)^2 |\Phi|^2$ for
a free field of mass $\mathfrak{m}$).
The corresponding energy-momentum tensor is
\be
    T_{\alpha\beta} = \nabla_{(\alpha} \bar\Phi \nabla_{\beta)} \Phi
    - \frac{1}{2} \left[ \nabla_\mu \bar\Phi \nabla^\mu \Phi + V(|\Phi|^2) \right]
    g_{\alpha\beta} .
\ee
Let us plug the above expression into (\ref{e:E_H}); using adapted 
coordinates $(t,r,\th,\phi)$ (cf. Sec.~\ref{s:adapted_coord}), we have
$\eta^\mu \nabla_\mu \Phi = \dert{\Phi}{t}$ and
$\ell^\mu \nabla_\mu \Phi = \dert{\Phi}{t} + \omega_H\dert{\Phi}{\phi}$.
In addition, 
$g_{\mu\nu} \eta^\mu \ell^\nu = 0$, since $\vw{\eta}$ is tangent 
to $\Hor$ and $\vw{\ell}$ is the normal to $\Hor$
(cf. Sec.~\ref{sub-section-Kerr-horizon}). Therefore, we get
\bea \label{e:E_H_scalar}
    \Delta E_H& =& 
    \int_{\Delta\Hor} 
    \left[ \der{\Phi}{t} \der{\bar\Phi}{t} + \frac{\omega_H}{2}
        \left( \der{\Phi}{t} \der{\bar\Phi}{\phi}
            + \der{\bar\Phi}{t} \der{\Phi}{\phi} \right) \right]\nonumber \\
 && \qquad \qquad\qquad \qquad\qquad    \times     \sqrt{q} \, \D t \, \D \th \, \D \phi .
\eea
Let us consider a rotating scalar field of the form 
\be
    \Phi(t,r,\th,\phi) = \Phi_0(r,\th) e^{i(\omega t - m\phi)}, 
\ee
where $\Phi_0(r,\th)$ is a real-valued function, $\omega$ is a constant 
and $m$ some integer. Then, (\ref{e:E_H_scalar}) becomes
\be
    \Delta E_H = 
    \int_{\Delta\Hor} \Phi_0^2 \omega(\omega-m\omega_H) \, 
    \sqrt{q} \, \D t \, \D \th \, \D \phi . 
\ee
In view of (\ref{e:Penrose_process}), 
we deduce immediately that a necessary and sufficient condition for a Penrose
process to occur is
\be \label{e:Penrose_process_scalar}
    0 < \omega < m\omega_H . 
\ee
In this context, the Penrose process is called \emph{super-radiance}
(see, e.g., \cite{Wald-1984} and \cite{LaszloR13}). 
{Condition (\ref{e:Penrose_process_scalar}) was obtained by 
Carter \cite{Carte79}  in the more general case of a
(not necessarily scalar) tensor field that is periodic in $t$
with period $2\pi/\omega$.}

\subsection{Perfect fluid}
\label{sub:pf}

Let us now consider a perfect fluid of 4-velocity $\vw{u}$, proper energy density 
$\varepsilon$ and pressure $p$. The corresponding energy-momentum tensor is
\be
\label{eq:Tpf}
    T_{\alpha\beta} = (\varepsilon + p) u_\alpha u_\beta + p g_{\alpha\beta} . 
\ee
Accordingly, and using $g_{\mu\nu} \eta^\mu \ell^\nu = 0$
as in Sec.~\ref{s:scalar_field}, formula (\ref{e:E_H}) becomes
\be \label{e:E_H_perfect_fluid}
    \Delta E_H 
        = \int_{\Delta\Hor}  (\varepsilon + p) \, \eta_\mu u^\mu  \,  \ell_\nu  u^\nu 
             \, \sqrt{q} \, \D t \, \D y^1 \, \D y^2  .
\ee
$\vw{\ell}$ being a future-directed null vector and $\vw{u}$ a future-directed timelike vector, we have
necessarily
\be \label{e:u_ell_neg}
    \ell_\nu  u^\nu  < 0 . 
\ee
According to (\ref{e:Penrose_process}), the Penrose process takes place if, and only if, $\Delta E_H < 0$. From (\ref{e:E_H_perfect_fluid}), (\ref{e:u_ell_neg}) and the assumption $\varepsilon+p \geq 0$ (the weak energy condition), we conclude that
for a perfect fluid, a necessary condition for the Penrose process to occur is
\be \label{e:cond_Penrose_fluid}
    { \eta_\mu u^\mu > 0  \ \ \mbox{in some part of $\Delta\Hor$} } .
\ee

We may have $\eta_\mu u^\mu > 0$ in some part of $\Delta\Hor$ only because $\vw{\eta}$ is there a spacelike vector
(for $\Hor$ is inside the ergoregion). 
Note that (\ref{e:u_ell_neg}) and (\ref{e:ell_eta_xi}) imply
\be
    \omega_H \xi_\mu u^\mu < - \eta_\mu u^\mu . 
\ee
Hence, in the parts of $\Delta\Hor$ where $\eta_\mu u^\mu > 0$, we have $\xi_\mu u^\mu < 0$. 
Therefore
for a perfect fluid, a necessary condition for the Penrose process to occur is
\be
    { \xi_\mu u^\mu < 0  \ \ \mbox{in some part of $\Delta\Hor$} } .
\ee
In other words, the fluid flow must have some azimuthal component
counterrotating with respect to the black hole 
in some part of $\Delta\Hor$. However, no physical process extracting black-hole rotational
energy through interaction with a perfect fluid is known.

In the special case of dust (fluid with $p=0$), the fluid lines are geodesics
and we recover from (\ref{e:cond_Penrose_fluid}) the single-particle condition 
$\Delta E_H < 0$, with $\Delta E_H$ given by (\ref{e:E_H_particle}).

\section{Electromagnetic fields}
\label{sect:em}

\subsection{General electromagnetic field}

Let us consider some electromagnetic field, described by the field 2-form $\w{F}$. 
For the moment we will deal with the most general case, i.e. that $\w{F}$ is not
necessarily stationary or axisymmetric. Of course this is possible only if $\w{F}$ is a passive field, 
i.e. does not contribute as a source to the Einstein equation, so that the spacetime metric remains stationary and axisymmetric. 

The electromagnetic energy-momentum tensor is given by the standard formula:
\be \label{e:T_EM}
    T_{\alpha\beta} = \frac{1}{\mu_0} \left( F_{\mu\alpha} F^\mu_{\ \, \beta}
	- \frac{1}{4} F_{\mu\nu} F^{\mu\nu} \; g_{\alpha\beta} \right) .
\ee
Accordingly, the integrand in formula (\ref{e:E_H}) for $\Delta E_H$ is
\[
    \w{T}(\vw{\eta},\vw{\ell}) = \frac{1}{\mu_0} \left( F_{\mu\rho} \eta^\rho F^\mu_{\ \, \sigma}
    \ell^\sigma 
	- \frac{1}{4} F_{\mu\nu} F^{\mu\nu} \; \vw{\eta}\cdot\vw{\ell} \right) . 
\]
Now, since $\vw{\eta}$ is tangent to $\Hor$ and $\vw{\ell}$ normal to $\Hor$, 
one has $\vw{\eta}\cdot\vw{\ell}= 0$. There remains then 
\be \label{e:Tem_eta_ell_FF}
    \mu_0  \w{T}(\vw{\eta},\vw{\ell}) = F_{\mu\rho} \eta^\rho F^\mu_{\ \, \sigma}
    \ell^\sigma . 
\ee
Let us introduce on $\Hor$ the ``pseudoelectric field'' 1-form (\cite{Carte73,Carte79,Damou79,Damou82})
\be \label{e:def_E}
    { \w{E} := \w{F}(., \vw{\ell}) }. 
\ee
If $\vw{\ell}$ were a unit timelike vector, $\w{E}$ would be a genuine electric field, namely the electric field measured by the observer whose 4-velocity is $\vw{\ell}$. But in the present case, 
$\vw{\ell}$ is a null vector, so that such a physical interpretation does not hold. 
$\w{E}$ is called a \emph{corotating electric field} in \cite{Carte73,Carte79}
because $\vw{\ell}$ is the corotating Killing vector on $\Hor$. 
Note that, \footnote{In this section, we are using index-free notations. In particular, the action of a 1-form on a vector is denoted by brackets, $\langle \w{E} , \vw{\ell} \rangle = E_\mu \ell^\mu$, and the scalar product of two vectors is denoted with a dot, 
$\vw{u}\cdot\vw{v} = g_{\mu\nu} u^\mu v^\nu = u_\nu v^\nu$.} thanks to the antisymmetry of $\w{F}$, 
\be \label{e:E_ell_zero}
    \langle \w{E} , \vw{\ell} \rangle = 0 . 
\ee
This implies that the vector $\vw{E}$ deduced from the 1-form $\w{E}$ by 
metric duality (i.e. the vector of components 
$E^\alpha = g^{\alpha\mu} E_\mu = F^\alpha_{\ \, \mu}\ell^\mu$)
is tangent to $\Hor$. Equation~(\ref{e:Tem_eta_ell_FF}) can be written as
\be \label{e:Tem_F_E_eta}
    { \mu_0  \w{T}(\vw{\eta},\vw{\ell}) = \w{F}(\vw{E},\vw{\eta}) }. 
\ee
Thanks to (\ref{e:ell_eta_xi}) and (\ref{e:def_E}), this expression can be recast as 
\bea
     \mu_0  \w{T}(\vw{\eta},\vw{\ell}) =\w{F}(\vw{E},\vw{\ell} - \omega_H \,\vw{\xi})
           & =& \w{F}(\vw{E},\vw{\ell}) - \omega_H  \w{F}(\vw{E},\vw{\xi})\nonumber \\
           & =& \langle \w{E},\vw{E} \rangle - \omega_H  \w{F}(\vw{E},\vw{\xi}) ,\nonumber
\eea
i.e. 
\be
\label{e:felectric}
    { \mu_0  \w{T}(\vw{\eta},\vw{\ell}) = \vw{E}\cdot\vw{E} - \omega_H  \w{F}(\vw{E},\vw{\xi}) 
        } . 
\ee
Given expression (\ref{e:E_H}) for $\Delta E_H$, we conclude that the necessary condition for the 
Penrose process to occur is 
\be \label{e:nec_cond_EM}
    { \omega_H \w{F}(\vw{E},\vw{\xi}) > \vw{E}\cdot\vw{E}  \ \ \mbox{in some part of $\Delta\Hor$} } . 
\ee
Note that since $\vw{E}$ is tangent to $\Hor$ [cf. (\ref{e:E_ell_zero})] and 
$\Hor$ is a null hypersurface, $\vw{E}$ is either a null vector or a spacelike one, 
so that in (\ref{e:nec_cond_EM}) one has always
\be
\label{e:Egt0}
    { \vw{E}\cdot\vw{E} \geq 0 } .
\ee
Equation (\ref{e:nec_cond_EM}) is the most general condition on any electromagnetic field configuration allowing
black-hole energy extraction through a Penrose process. Obviously, for $\omega_H=0$ there is no energy extraction.

\subsection{Stationary and axisymmetric electromagnetic field}

In this section, we assume that the electromagnetic field obeys the spacetime symmetries, 
which is expressed by
\be
    \Lie_{\vw{\eta}} \w{F} = 0 \qquad\mbox{and}\qquad \Lie_{\vw{\xi}} \w{F} = 0 , 
\ee
where $\Lie_{\vw{v}}$ stands for the Lie derivative along the vector field $\vw{v}$. 
Then it can be shown (see e.g. \cite{GourgMUE11} for details) that
$\w{F}$ is entirely determined by three scalar fields $\Phi$, $\Psi$, and $I$ such that
\bea 
    & & \w{F}(.,\vw{\eta}) = \dd\Phi \label{e:def_Phi} \\
    & & \w{F}(.,\vw{\xi}) = \dd\Psi  \label{e:def_Psi}  \\
    & & {}^\star\w{F}(\vw{\eta},\vw{\xi}) = I , \label{e:def_I}
\eea
where $\dd$ is the exterior derivative operator (reducing to the gradient for a scalar field
such as $\Phi$ or $\Psi$) and ${}^\star\w{F}$ stands for the Hodge dual of $\w{F}$.
Note that, being defined solely from $\w{F}$ and the Killing fields $\vw{\eta}$ and $\vw{\xi}$, 
$\Phi$,  $\Psi$, and $I$ are gauge-independent quantities. Introducing an electromagnetic potential 
1-form $\w{A}$ such that $\w{F} = \dd\w{A}$, one may use the 
standard electromagnetic gauge freedom to choose $\w{A}$ so that
\be
    \Phi = \langle \w{A}, \vw{\eta} \rangle = A_t
    \qquad\mbox{and}\qquad
    \Psi = \langle \w{A}, \vw{\xi} \rangle = A_\varphi . 
\ee
In addition to (\ref{e:def_Phi})-(\ref{e:def_I}), one has (see e.g. \cite{GourgMUE11})
$\w{F}(\vw{\eta},\vw{\xi}) = 0$
and 
\be \label{Phi_Psi_stax}
    \Lie_{\vw{\eta}} \Phi = \Lie_{\vw{\xi}} \Phi = 0 
    \qquad\mbox{and}\qquad
     \Lie_{\vw{\eta}} \Psi = \Lie_{\vw{\xi}} \Psi = 0 ,     
\ee
which means that the scalar fields $\Phi$ and $\Psi$ obey the two spacetime symmetries. 

From the definition (\ref{e:def_E}) and expression (\ref{e:ell_eta_xi}) of $\vw{\ell}$, 
the corotating pseudoelectric field $\w{E}$ is
\[
    \w{E} = \w{F}(., \vw{\ell}) =  \w{F}(., \vw{\eta}) + \omega_H \w{F} (., \vw{\xi}) 
        = \dd\Phi + \omega_H \dd\Psi , 
\]
where the last equality follows from (\ref{e:def_Phi}) and (\ref{e:def_Psi}). 
Since $\omega_H$ is constant, we conclude that the 1-form $\w{E}$ is a pure gradient:
\be \label{e:E_dPhi_omH_Psi}
    {\w{E} = \dd(\Phi + \omega_H \Psi) } . 
\ee
\emph{Remark:} If the electromagnetic field is not passive, i.e. if it contributes  
significantly to the spacetime metric via the Einstein equation, 
then $\w{T}(\vw{\ell},\vw{\ell})$ must vanish in order for the black hole to be in equilibrium
(otherwise it would generate some horizon expansion, via the Raychaudhuri equation;
see, e.g., \cite{Carte73}. 
Since by (\ref{e:T_EM}), $\w{T}(\vw{\ell},\vw{\ell}) = \mu_0^{-1} \vw{E}\cdot\vw{E}$, 
this implies that $\vw{E}$ is a null vector. Being tangent to $\Hor$, the only possibility
is to have $\vw{E}$ collinear to $\vw{\ell}$:  $\vw{E} = f \vw{\ell}$. Then for any
vector $\vw{v}$ tangent to $\Hor$, one has $\vw{v}\cdot\vw{E} = 0$. 
In view of (\ref{e:E_dPhi_omH_Psi}), we get the remarkable result that
{\cite{Carte73}}
\be 
\label{e:const_pot}
    \Phi + \omega_H \Psi \mbox{\ is constant over\ }\Hor.
\ee

Returning to the case of passive fields we notice that thanks to (\ref{e:def_Phi}), the $\Delta E_H$ integrand (\ref{e:Tem_F_E_eta}) becomes
\be
\label{e:edphi}
    \mu_0  \w{T}(\vw{\eta},\vw{\ell}) = \vw{E} \cdot\vw{\nabla} \Phi . 
\ee
In a similar way, from (\ref{e:def_Psi}) one deduces that the $\Delta J_H$ integrand $\mu_0  \w{T}(\vw{\xi},\vw{\ell}) = \w{F}(\vw{E},\vw{\xi})$ takes the form of
\be
\label{e:edpsi}
    \mu_0  \w{T}(\vw{\xi},\vw{\ell}) = \vw{E} \cdot\vw{\nabla} \Psi . 
\ee 
In view of (\ref{e:E_dPhi_omH_Psi}), we get
\be \label{e:T_eta_ell_EM_stax}
   { \mu_0  \w{T}(\vw{\eta},\vw{\ell}) = \vw{\nabla} \Phi \cdot \vw{\nabla} (\Phi + \omega_H \Psi) } . 
\ee

\subsection{Force-free stationary and axisymmetric field (Blandford-Znajek)}

Let us assume that the electromagnetic field is force free, in addition of
being stationary and axisymmetric:
\be
    \w{F}(\vw{j}, .) = 0 , 
\ee
where $\vw{j}$ is the electric 4-current. 
In particular,  $\w{F}(\vw{j},\vw{\eta}) = 0$ and $\w{F}(\vw{j},\vw{\xi}) = 0$.
From (\ref{e:def_Phi}) and (\ref{e:def_Psi}), it follows immediately that
\be \label{e:jdPhi_zero}
    \vw{j} \cdot\vw{\nabla} \Phi = 0 
    \qquad\mbox{and}\qquad
    \vw{j} \cdot\vw{\nabla} \Psi = 0 . 
\ee
Taking into account that $\Phi$ and $\Psi$ are stationary and axisymmetric 
[cf. (\ref{Phi_Psi_stax})], we may rewrite (\ref{e:jdPhi_zero}) in a coordinate system 
$(t,r,\theta,\varphi)$ adapted to stationarity and axisymmetry as 
\[
    j^r \der{\Phi}{r} + j^\theta \der{\Phi}{\theta} = 0 
    \qquad\mbox{and}\qquad
     j^r \der{\Psi}{r} + j^\theta \der{\Psi}{\theta} = 0 . 
\]
We deduce that, generically, there exists a function $\omega = \omega(\Psi)$ such that
\be
\label{e:dephidepsi}
    \dd \Phi = -\omega(\Psi) \dd\Psi . 
\ee
Equation (\ref{e:T_eta_ell_EM_stax}) becomes then 
\be \label{e:T_eta_ell_BZ}
   {  \mu_0  \w{T}(\vw{\eta},\vw{\ell}) = \omega(\Psi) \left( 
    \omega(\Psi) - \omega_H \right) \, \vw{\nabla}\Psi \cdot\vw{\nabla} \Psi } . 
\ee
Notice also that from (\ref{e:edphi}),  (\ref{e:edpsi}) and (\ref{e:dephidepsi}) it follows that for an axisymmetric, stationary and force-free field
\be
\label{e:eomegj}
\Delta E_H=\omega(\Psi) \Delta J_H.
\ee
Now, we have
\[
    \vw{\ell}\cdot  \vw{\nabla}\Psi = \vw{\eta} \cdot  \vw{\nabla}\Psi
        + \omega_H \vw{\xi} \cdot  \vw{\nabla}\Psi
            = \underbrace{\Lie_{\vw{\eta}} \Psi}_{0} 
            + \omega_H \underbrace{\Lie_{\vw{\xi}} \Psi}_{0} = 0 . 
\]
This means that the vector $\vw{\nabla}\Psi$ is tangent to $\Hor$. 
Since the latter is a null hypersurface, it follows that $\vw{\nabla}\Psi$ is either null or spacelike. 
Therefore, on $\Hor$,
\be
    \vw{\nabla}\Psi \cdot\vw{\nabla} \Psi \geq 0 . 
\ee
Accordingly (\ref{e:T_eta_ell_BZ}) yields
\[
    \w{T}(\vw{\eta},\vw{\ell}) < 0 \iff 
    \left\{ \begin{array}{l}
        \omega(\Psi) \left( 
    \omega(\Psi) - \omega_H \right) < 0\\
        \vw{\nabla}\Psi \cdot\vw{\nabla} \Psi \not= 0 .
        \end{array}\right.
\]
i.e. 
\be
    { \w{T}(\vw{\eta},\vw{\ell}) < 0 \iff 
    \left\{ \begin{array}{l}
        0 < \omega(\Psi) < \omega_H \\
        \vw{\nabla}\Psi \cdot\vw{\nabla} \Psi \not= 0
        \end{array}\right.} . 
\ee
We recover the result (4.6) of Blandford and Znajek's article \cite{Blandford-1977}. 
In view of (\ref{e:E_H}) and (\ref{e:Penrose_process}), we may conclude the following: 
\begin{quote}
For a stationary and axisymmetric force-free electromagnetic field, a necessary condition for the 
Penrose process to occur is 
\be
\label{e:bzomega}
    { 0 < \omega(\Psi) < \omega_H\ \ \mbox{in some part of $\Delta\Hor$} } .
\ee
\end{quote}
In particular, for a nonrotating black hole ($\omega_H = 0$), no Penrose process can occur.
The condition (\ref{e:bzomega}) can be compared to the condition 
(\ref{e:Penrose_process_scalar}) for a scalar field. 

\section{Simulations of electromagnetic extraction of black-hole rotational energy}
\label{sect:mad}

Until recently, the relevance of the Blanford-Znajek process to observed
high energy phenomena such as relativistic jets has been hotly debated
and the efficiency of this mechanism put in doubt
(see, e.g., \cite{Ghosh-1997,Livio-1999}). Providing jet production efficiencies of less than $\sim 20\%$, general relativistic magnetohydrodynamic (GRMHD) simulations were not of much help in ending the controversy. Only recently a new physical setup of GRMHD simulations (\cite{Tchekhovskoy-2011,McKinney-2012}) produced the first clear evidence of net energy extraction by magnetized accretion onto a spinning black hole.
These simulations were carried out with general relativistic MHD code
HARM \citep{Gammie03} with recent improvements \citep{mb09,Tchekhovskoy-2011}.

\subsection{The framework}
\label{sect:framework}

The BZ efficiency can be defined as BZ power normalized by $\dot Mc^2$:
\be
\label{e:BZ_eff}
\eta_{\rm BZ}=\frac{\left[P_{\rm BZ}\right]_t}{\left[\dot
    M\right]_tc^2}=\frac{\kappa}{4\pi  c}\left[\phi^2_{BH}\right]_t
\left(\frac{\omega_H r_g}{c}\right)^2 f(\omega_H)
\ee
where $\dot M$ is the accretion rate; $\left[.\, .\, . \, \right]_t$ designates the time average; $\kappa\approx0.05$ depends weakly on the magnetic field geometry, $\phi_{BH}^2={\Phi}^2_{BH}/\dot M r_g^2 c$, $\Phi_{BH}$ being the magnetic flux through the black-hole surface; $f(\omega_H) \approx 0.77$  for $a_*=1$, where $a_*=J/m^2$ \citep{Tchekhovskoy-2012}; $r_g=Gm/c^2$ is black-hole gravitational radius.

The efficiency $\eta_{\rm BZ}$ depends on spin and the magnetic flux
on the black hole. The spin  is limited by $a_*< 1.0$ ($\omega_H <
c/r_S$; where $r_S=2Gm/c^2$); the magnetic flux is limited by two
factors. (1) How much of it can be pushed on to the black hole. (2) How
much of it can be accumulated by diffusion through the accretion
flow. In an MHD turbulent disk, accumulation of dynamically-important
magnetic field is possible only if it is not geometrically thin,
i.e. only if $h/r \sim 1$ \citep{Lubow-94}. \citet{Tchekhovskoy-2011}
considered ``slim" disks ($h/r\sim0.3$) in which initially poloidal
magnetic fields are accumulated at the black hole until they obstruct
the accretion and lead to the formation of a so-called magnetically
arrested disk (\cite{Igu2003,Narayan-2003}). In such a configuration $\phi_{BH} \sim 40$ for $a_*=0.99$, leading to $\eta_{\rm BZ} > 100\%$, i.e., to {\sl net} energy extraction from 
a rotating black hole.

This result, as well as subsequent simulations of various MAD\footnote{These were also called magnetically choked accretion flows by \citet{McKinney-2012}.} configurations \citep[][]{McKinney-2012}, leaves little doubt that the Blandford-Znajek mechanism can play a fundamental role in launching of (at least some) relativistic jets from the vicinity of black-hole surfaces. This conclusion is supported
by observational evidence of the role of spin {\sl and} accumulated magnetic flux in launching of relativistic jets both in microquasars and active galactic nuclei (see, e.g.,\cite{Narayan-2012,MNS13,NMT13,siksta13,sb13}).

In the previous section we obtained several conditions for the occurrence of a Penrose process in the presence of electromagnetic fields. All these criteria follow from the fundamental requirement $\Delta E_H <0$. The most
general criterion applies to any electromagnetic field configuration: from the definition (\ref{e:E_H}) and the general condition  (\ref{e:Penrose_process}) we deduced a specific (necessary) condition (\ref{e:nec_cond_EM}) for the electromagnetic fields on the horizon. 
We then showed that in the case of stationary and axisymmetric force-free fields the condition (\ref{e:Penrose_process})
is equivalent to the \citet{Blandford-1977} condition on the angular velocity of the magnetic field lines. In this section we will apply
these conditions to the results GRMHD simulations of magnetized jets we have discussed above. The aim of this exercise is twofold. First, using rigorous general-relativistic criteria we will confirm that the MAD BZ mechanism is indeed a Penrose process as surmised by \citet{Tchekhovskoy-2011}. Second, our Penrose-process conditions can be used as a diagnostic tool to test the physical and mathematical consistency of numerical calculations reputed to represent the Blandford-Znajek/Penrose process.

In dealing with results of numerical simulations, 
we will adopt the 3+1 Kerr coordinates $(t,r,\th,\phi)$ described in Appendix~\ref{ap:Kerr}, 
which are adapted coordinates in the sense defined in Sec.~\ref{s:adapted_coord}. 
The energy captured by the black hole over $\Delta\Hor$ is given by 
(\ref{e:E_H_adapted}). Since for the 3+1 Kerr coordinates, 
$\sqrt{-g} = (r^2 + a^2 \cos^2\th)\sin\th$ [cf. (\ref{e:det_g_Kerr})], we get
\be \label{e:ehindex}
    \Delta E_H = \int_{\Delta\Hor} {\dot e}_H \, 
    (r_H^2 + a^2 \cos^2\th)\sin\th \, \D t \, \D \theta 
    \, \D \phi , 
\ee
where we have defined
\be \label{e:def_dot_e_H}
    {\dot e}_H := - \left. P^r \right| _{\Hor} 
        =  \left. T^r_{\ \, t} \right| _{\Hor} . 
\ee
As a check of (\ref{e:ehindex}), we may recover it from the last integral 
in Eq. (\ref{e:E_H}), noticing that $\eta^\mu = (1,0,0,0)$,
$\ell_r = (r_H^2 + a^2 \cos^2\th) / (2m r_H)$, 
and $\sqrt{q} = 2 m r_H \sin\theta$ [cf. (\ref{e:sqrt_det_q}) in Appendix~\ref{ap:Kerr}].

A formula analogous to (\ref{e:ehindex}), with 
${\dot e}_H$ replaced by $-T^r_{\ \, \phi}$, 
gives $\Delta J_H$ [cf. (\ref{e:J_H_adapted})]; accordingly, we define
\be \label{e:def_dot_j_H}
    {\dot \jmath}_H := - \left. M^r \right| _{\Hor} 
        =  - \left. T^r_{\ \, \phi} \right| _{\Hor} . 
\ee

Since,  as discussed in Sec.\,\ref{subsect: applications}, in numerical simulations one assumes stationarity, and $\Sigma_2$ is deduced from $\Sigma_1$ by time translation, one
must have $E_2=E_1$ (see Fig. \ref{f:Econs}). 
Therefore, to test the Penrose-process condition (\ref{e:Penrose_process}) and (\ref{e:omegJ_EH}) and show the details of the BZ mechanism, we found it convenient to use the energy and angular-momentum \emph{flux densities} 
$\dot e_H(t,\theta,\phi)$ and $\dot {\jmath}_H(t,\theta,\phi)$ defined by (\ref{e:def_dot_e_H}) and (\ref{e:def_dot_j_H}),
and plot their ($t-$ and $\phi$-averaged) longitudinal distribution on the $t$-constant 2-surface ${\Sps_t}$ (the black hole's surface; see Sec. \ref{sub-section-Kerr-horizon}) on $\Hor$. 

In the MAD simulations the energy-momentum tensor is the sum of the perfect fluid (\ref{eq:Tpf}) and the electromagnetic (\ref{e:T_EM}) tensors:
\[
T_{\mu\nu}= T^{\rm( MA)}_{\mu\nu} + T^{\rm (EM)}_{\mu\nu}.
\]
Consequently we define  $\dot e_{\rm MA}:=T^{{\rm( MA)}\,r}_{\qquad  t}$ and $\dot {\jmath}_{\rm MA}:=- T^{{\rm( MA)}\,r}_{\qquad \phi}$; $\dot e_{\rm EM}$ and $\dot {\jmath}_{\rm EM}$ are defined in an analogous way through the electromagnetic energy-momentum tensor.
In the simulation of force-free fields $\dot e_{\rm MA}=0$.
\begin{figure}
\centerline{
\includegraphics[width=0.4\textwidth]{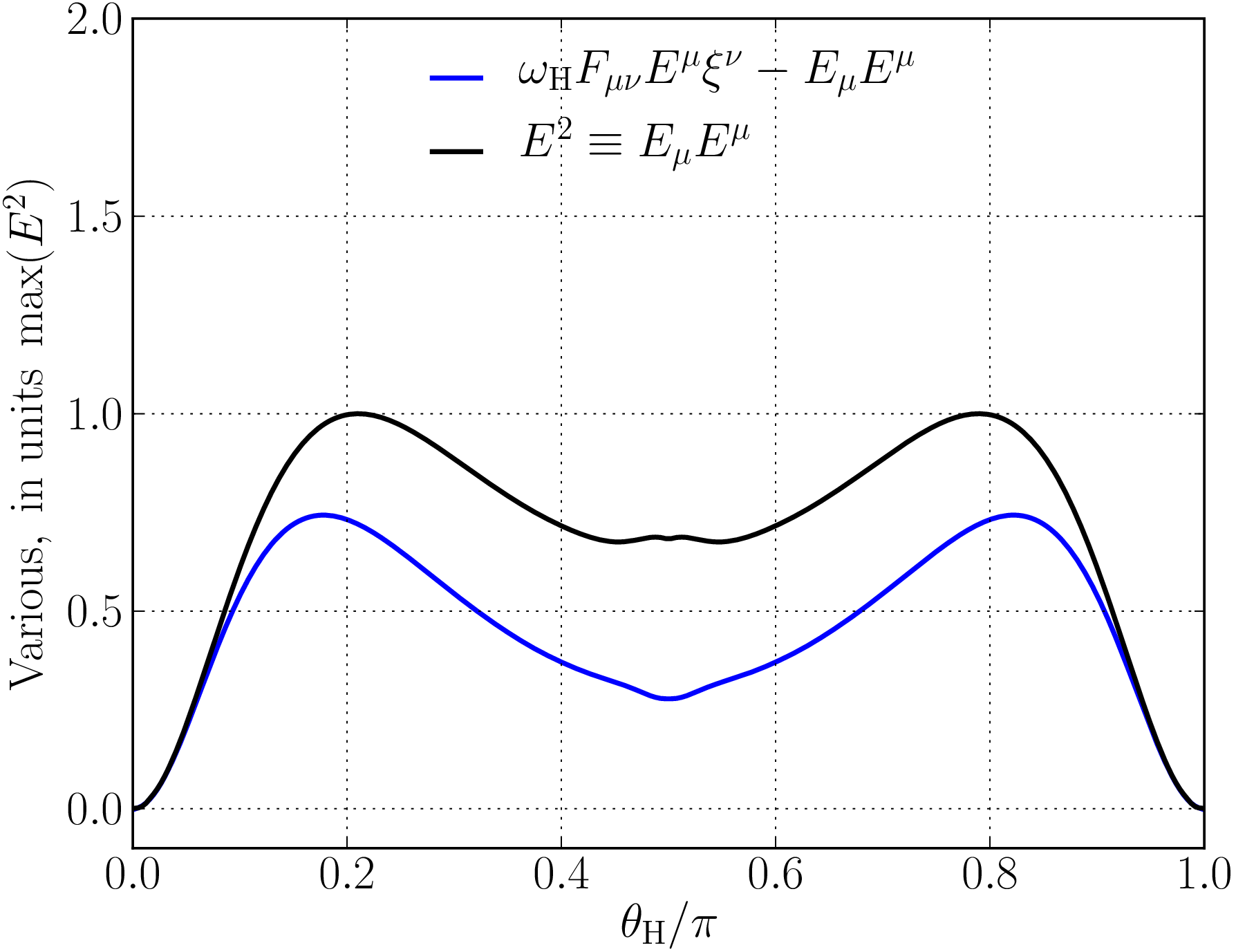}}
\caption{\small
Values of  $\omega_H F_{\mu\nu} E^\mu \xi^\nu - E_\mu E^\mu$ and $E_\mu E^\mu$ plotted as a function of $\theta$ for a stationary, axisymmetric and force-free field with $a_*=0.99$. The necessary condition: (\ref{e:penrose_nc}) for the occurrence of a Penrose process is satisfied over all  ${\Sps_t}$ (except the poles).}
\label{f:Electric}
\end{figure}
\begin{figure}
  \centerline{
  \includegraphics[width=0.4\textwidth]{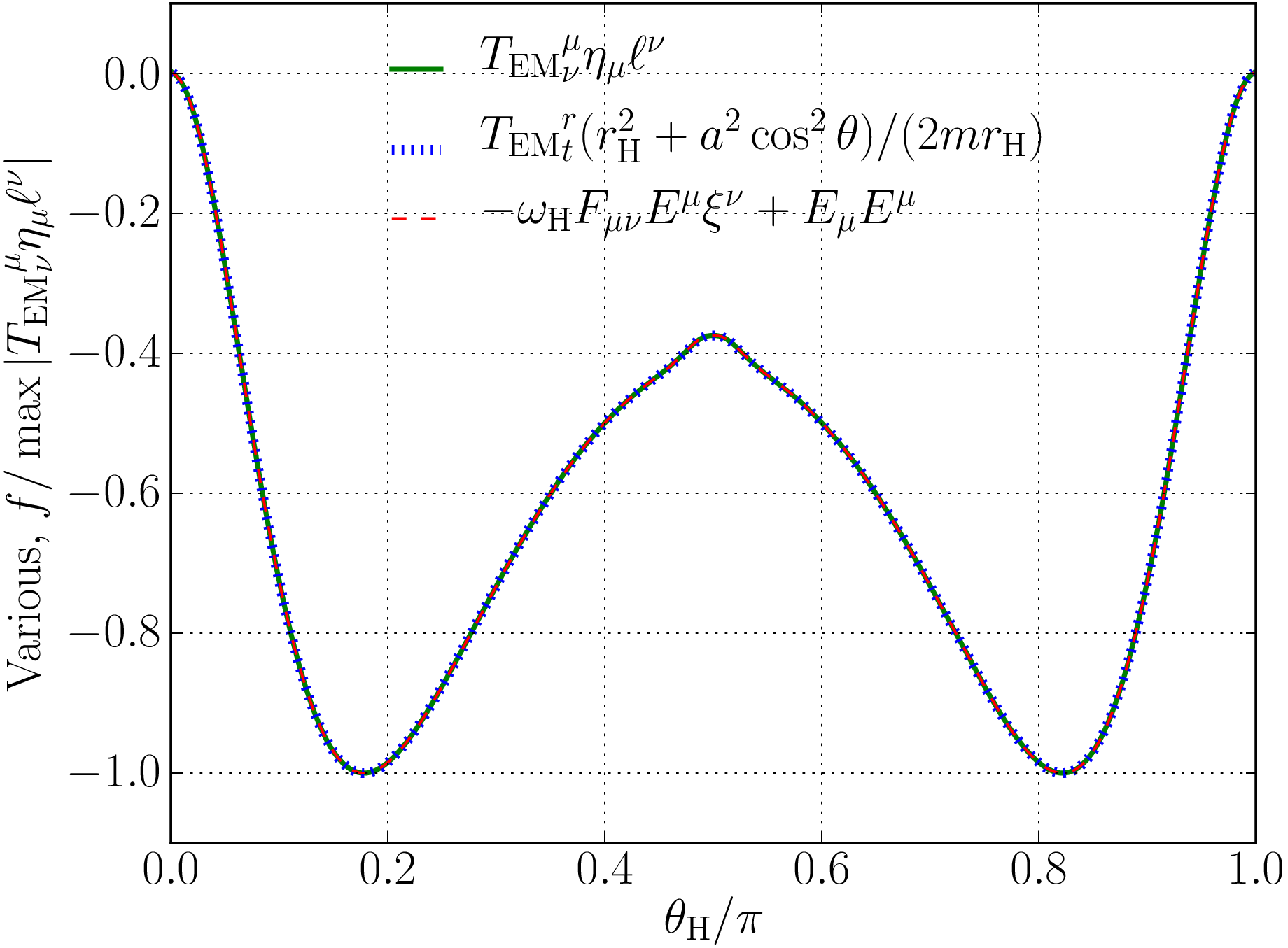}}
  \caption{\small Comparison of $-\omega_H F_{\mu\nu} E^\mu \xi^\nu + E_\mu E^\mu$ with the integrands of (\ref{e:ehindex}) for the same field configuration and black-hole spin as in Fig.~\ref{f:Electric}.  }
  \label{f:condition}
 \end{figure}

The pseudoelectric field (\ref{e:def_E}) is $E_\alpha = F_{\alpha\mu} \ell^\mu$.
Therefore in the index notation, the general necessary condition (\ref{e:nec_cond_EM}) for the Penrose process to occur takes the form
\be
\label{e:penrose_nc}
\omega_H F_{\mu\nu} E^\mu \xi^\nu - E_\mu E^\mu\vert_{\Hor}>0.
\ee
In the case of MAD simulations, which are intrinsically time variable, we run the simulations long enough to achieve quasisteady state in which all quantities fluctuate about their mean values, so we use the time average of the left-hand side 
in (\ref{e:penrose_nc}).

\subsection{Force-free stationary electromagnetic field}
\label{subsect:ff}

As a warm-up, we present the results of simulations of  black-hole rotational energy extraction by a 
force-free electromagnetic field. As illustration, we consider the simple case of a paraboloidal magnetic field for an $a_*=0.99$ black hole. The field configuration corresponds to the $\nu=1$ case of \citet{Tchekhovskoy-2010}, where additional
information about the setup of the problem can be found. We have chosen a paraboloidal field in preference to a monopole because of the similarity of results with those of MAD simulations discussed later.
\begin{figure}
  \centerline{
  \includegraphics[width=0.4\textwidth]{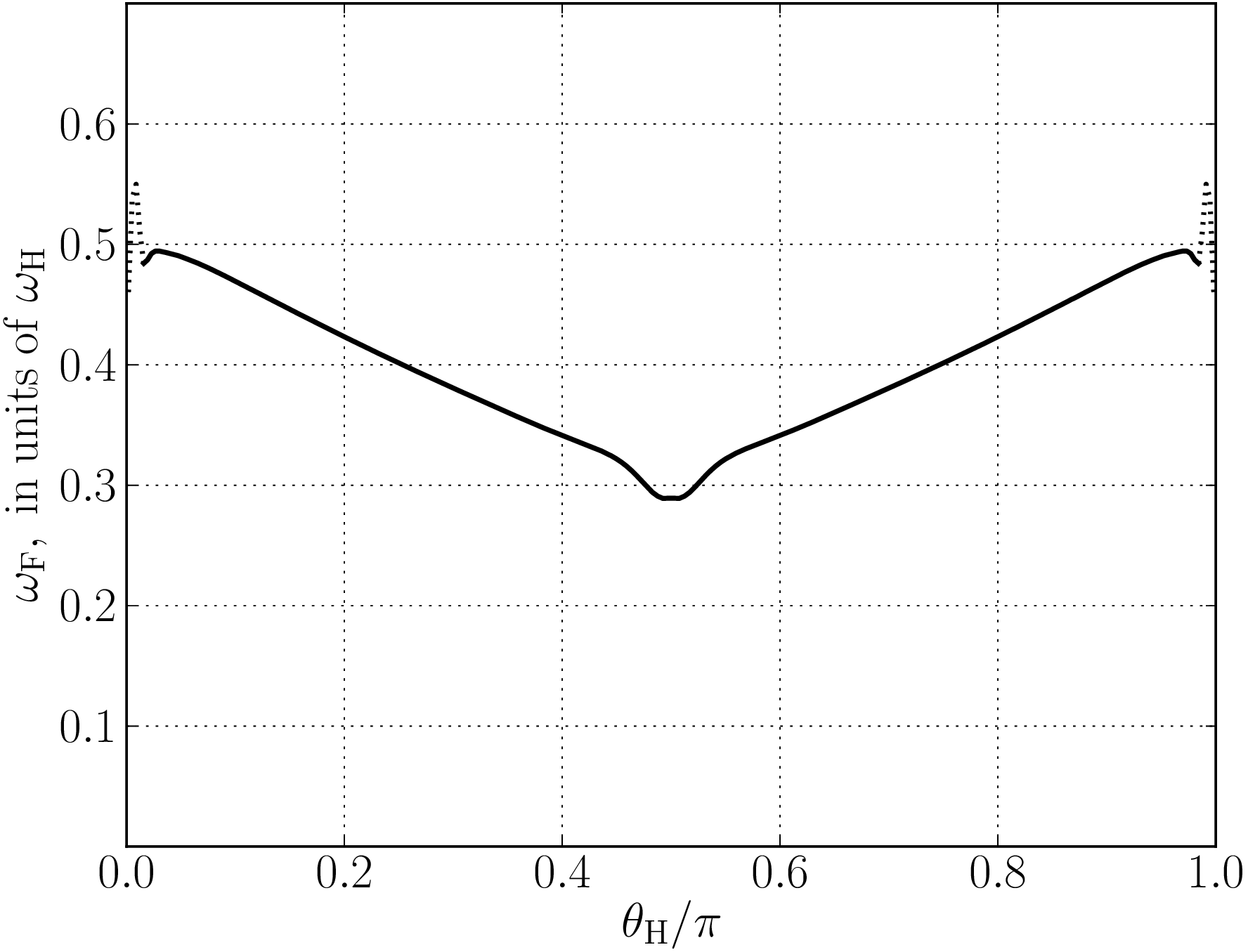}}
  \caption{\small
$\omega_F/\omega_H$ plotted as function of $\theta$ for a stationary, axisymmetric force-free  paraboloidal magnetic field;  $a_*=0.99$. The condition (\ref{e:bzomega}) for the occurrence of a Penrose process is satisfied over the entire black-hole 2-surface ${\Sps_t}$.}
 \label{f:omegaf}
\end{figure}
\begin{figure}
  \centerline{
  \includegraphics[width=0.4\textwidth]{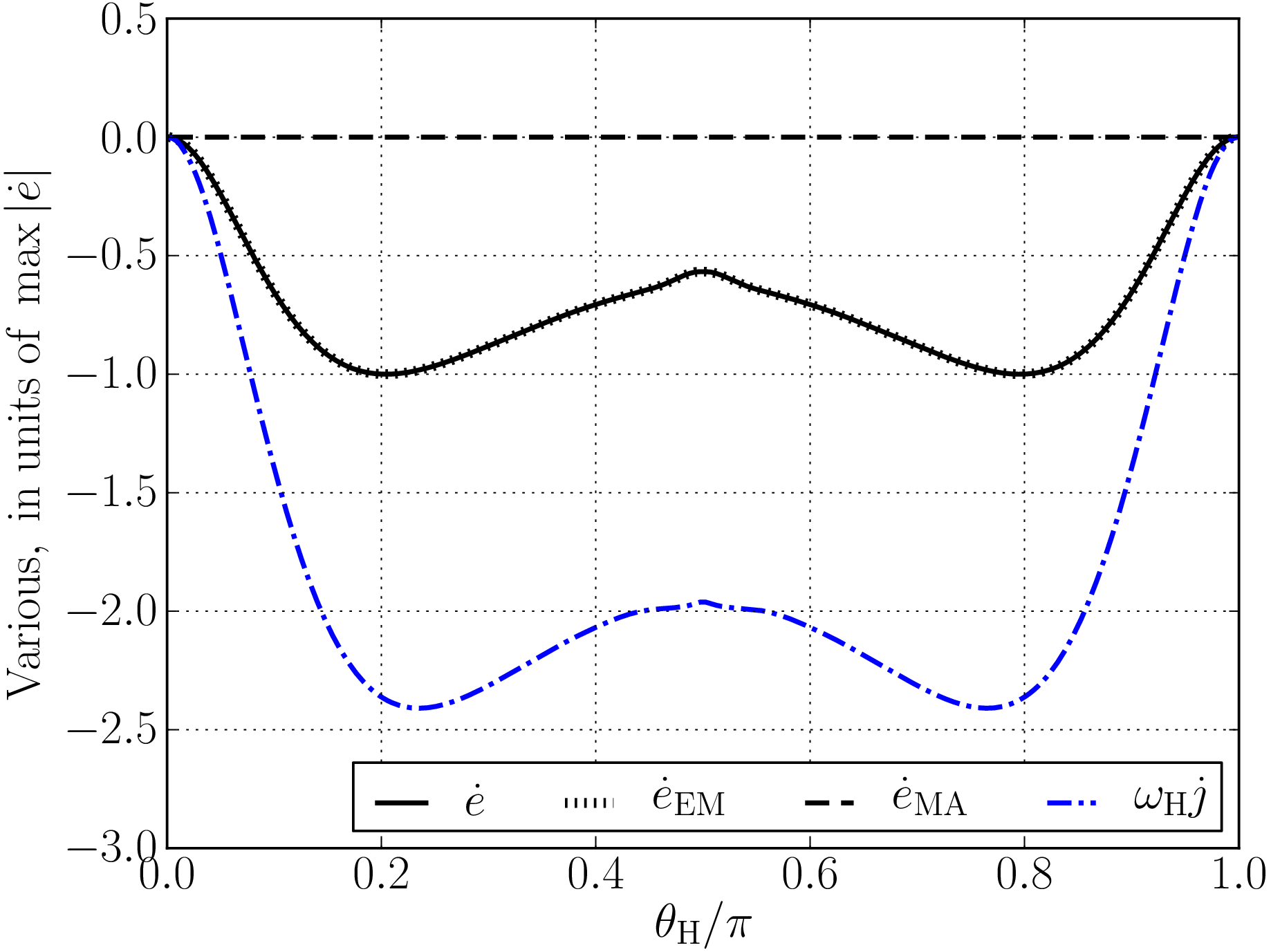}}
  \caption{\small Values of energy and angular-momentum density fluxes on the black-hole surface as function of  $\theta$ for a force-free field and $a_*=0.99$. In this case $\dot e = \dot e_{\rm EM}$. $\dot e$ is everywhere negative on ${\Sps_t}$ in agreement with the Penrose-process condition (\ref{e:Penrose_process}); the same is true by construction of   $\dot {\jmath}$ and (\ref{e:Penrose_am}) is obviously satisfied.
  }
  \label{f:eminusomegal}
\end{figure} 

First, in Fig.~\ref{f:Electric} we present the results of testing the general condition (\ref{e:penrose_nc}). It is satisfied on the whole of  the black-hole surface ${\Sps_t}$. Also (\ref{e:Egt0}) is satisfied which confirms that the simulations correctly reproduce the spacetime structure near and at the horizon. Since condition (\ref{e:penrose_nc}) follows from the requirement of negative energy on the horizon we checked the consistency of the numerical scheme by comparing the expression $- \omega_H F_{\mu\nu} E^\mu \xi^\nu - E_\mu E^\mu$ with two forms of the integrand in (\ref{e:ehindex}). As expected the values of the two expressions are identical (see Fig. \ref{f:condition}).

The force-free BZ condition (\ref{e:bzomega}) is satisfied everywhere on the black hole's surface (Fig. \ref{f:omegaf}). Since in a force-free field $\dot e_H=\omega_H \dot {\jmath}_H$
[cf. (\ref{e:eomegj})] the Penrose-process condition (\ref{e:Penrose_am}) follows directly from  $\Delta E_H <0$  [Eq.~(\ref{e:Penrose_process})]; see Fig. \ref{f:eminusomegal}.

Since it satisfies the required conditions on the horizon, the BZ
mechanism described by numerical simulations of the interaction of a
force-free field with a spinning black hole is a Penrose process.

\subsection{Magnetically arrested disks}
\label{subsect:MAD}

Before discussing the results of GRMHD MAD simulations in the context of the BZ/Penrose mechanism, we have first to present the underlying assumptions in more detail.

The simulations are performed in a ``box" of {\sl finite} size delimited by $\Delta \Hor$ and $\Sigma_{\rm ext}$ in space and $\Sigma_1$ and 
$\Sigma_2$ in time.

It is supposed that $\Sigma_{\rm ext}$ is located at some reasonably large radius ($\gta 30 r_g$), which is far from the horizon but still well inside the converged volume of the simulation.
One also assumes that the times $t_1$ and $t_2$ corresponding, respectively, to $\Sigma_1$ and $\Sigma_2$ are sufficiently far apart so that time averages are well defined and the system is in a steady state during this time.
In a steady state $E_2=E_1$; i.e., the energy contained inside the volume defined by the boundaries $\Delta \Hor$ and $\Sigma_{\rm ext}$ is independent of time.

Simulation shows that $\Delta E_{\rm ext}>0$, i.e., there is a net flow of energy out of the system. From energy conservation (\ref{e:ener_cons}) one should therefore have 
$\Delta E_H <0$ on some part of $\Delta \Hor$.
Below we will show that stationary MAD models of energy extraction from a spinning black hole satisfy this condition and are an electromagnetic realization of a Penrose process.
\begin{figure}
\centering
\centering
\includegraphics[width=0.4\textwidth]{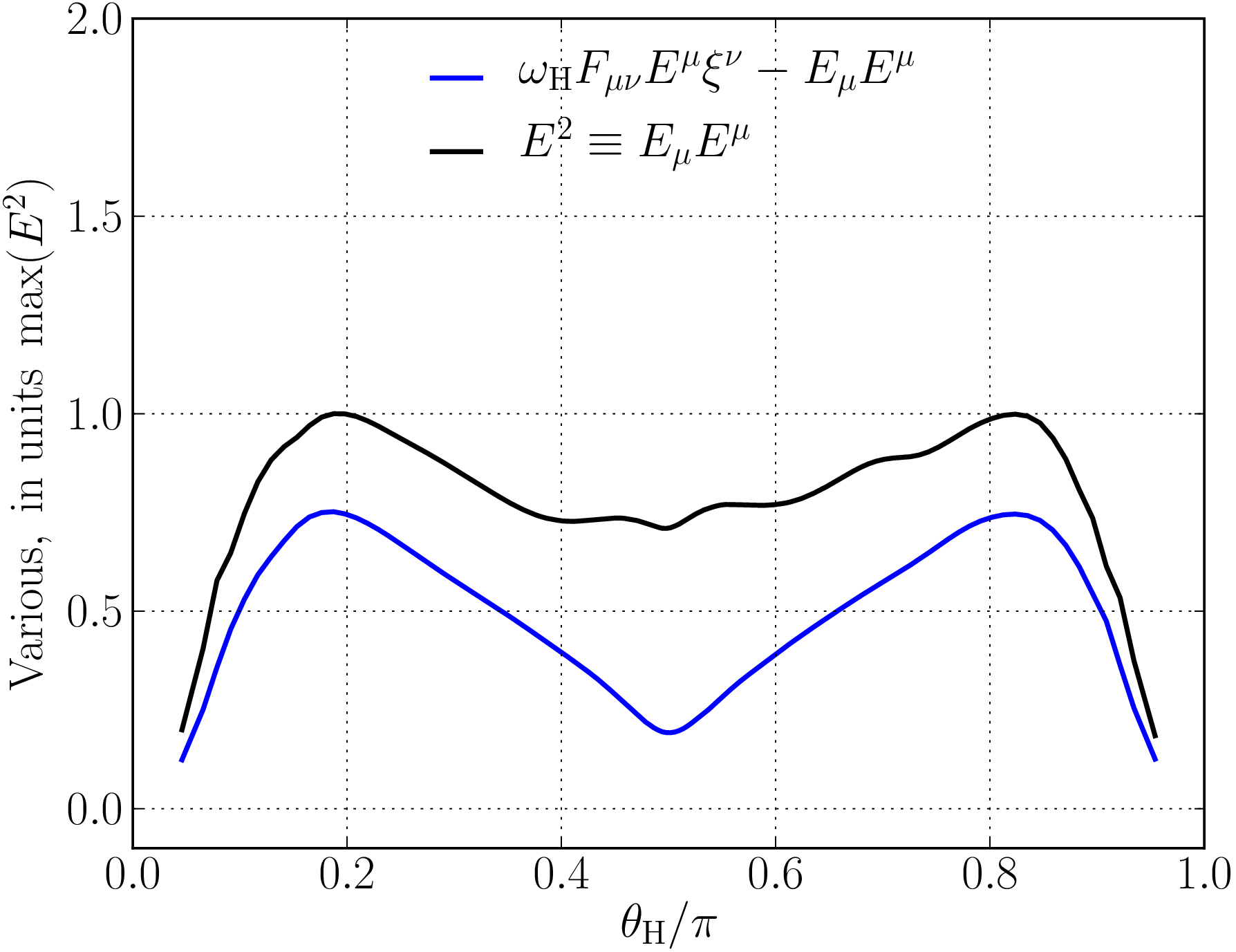}
\caption{  \small
Same as in Fig. \ref{f:Electric} but for a MAD simulation with $a_*=0.99$. Here the time- and $\phi$-averaged quantities are used: $\omega_H \langle F_{\mu\nu} E^\mu\,\rangle \xi^\nu - \langle E_\mu E^\mu\, \rangle$ and $\langle E_\mu E^\mu \rangle$. The necessary condition (\ref{e:penrose_nc}) for the occurrence of a Penrose process is satisfied over all ${\Sps_t}$.}
\label{f:Emad}
\end{figure}
\begin{figure}
\centering
  \includegraphics[width=0.4\textwidth]{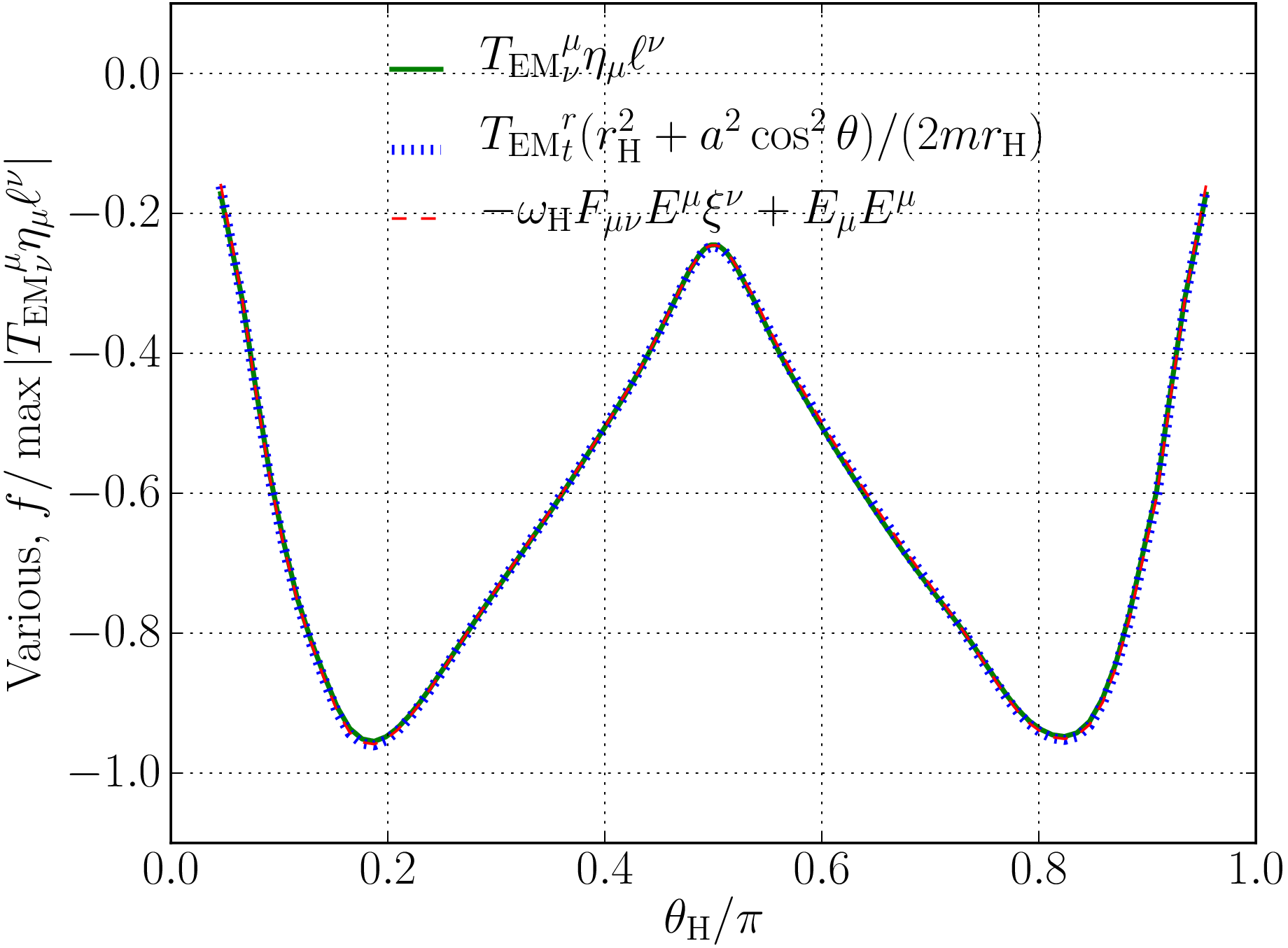}
  \caption{\small Comparison of $-\omega_H \langle F_{\mu\nu} E^\mu \rangle \xi^\nu + \langle E_\mu E^\mu\rangle$ with the integrands of (\ref{e:ehindex}) for the same field configuration and black-hole spin as in Fig. \ref{f:Emad}.}
  \label{f:madcondition}
\end{figure}

We will use the results of the model A0.99N100 of \citet{McKinney-2012}. In this model the initial magnetic field is poloidal, $a_*=0.99$, and the disk is moderately thick: the half-thickness $h$ satisfies $h/r \sim 0.3$ at $R_{\rm ext}=30 r_g$ and $h/r\lta 0.1$ at the black-hole surface.

We will first examine if the MAD simulations satisfy the Penrose-process conditions (\ref{e:penrose_nc}), (\ref{e:Penrose_process}) and (\ref{e:Penrose_am}). As for the force-free fields, we start with checking condition (\ref{e:penrose_nc}) for the electromagnetic fields on the black-hole surface. As shown in Fig. \ref{f:Emad} 
$\omega_H \langle F_{\mu\nu} E^\mu\,\rangle \xi^\nu - \langle E_\mu E^\mu\, \rangle >0 $ everywhere on the black-hole surface, which implies that the electromagnetic energy is negative everywhere on $\Delta\Hor$. Indeed, as shown in Fig. \ref{f:madcondition} the electromagnetic energy density $T^{EM}_{\mu\nu}\eta^\mu\ell^\nu$ is everywhere negative on the black-hole surface. 
\begin{figure}
\centerline{
\includegraphics[height=0.35\textwidth]{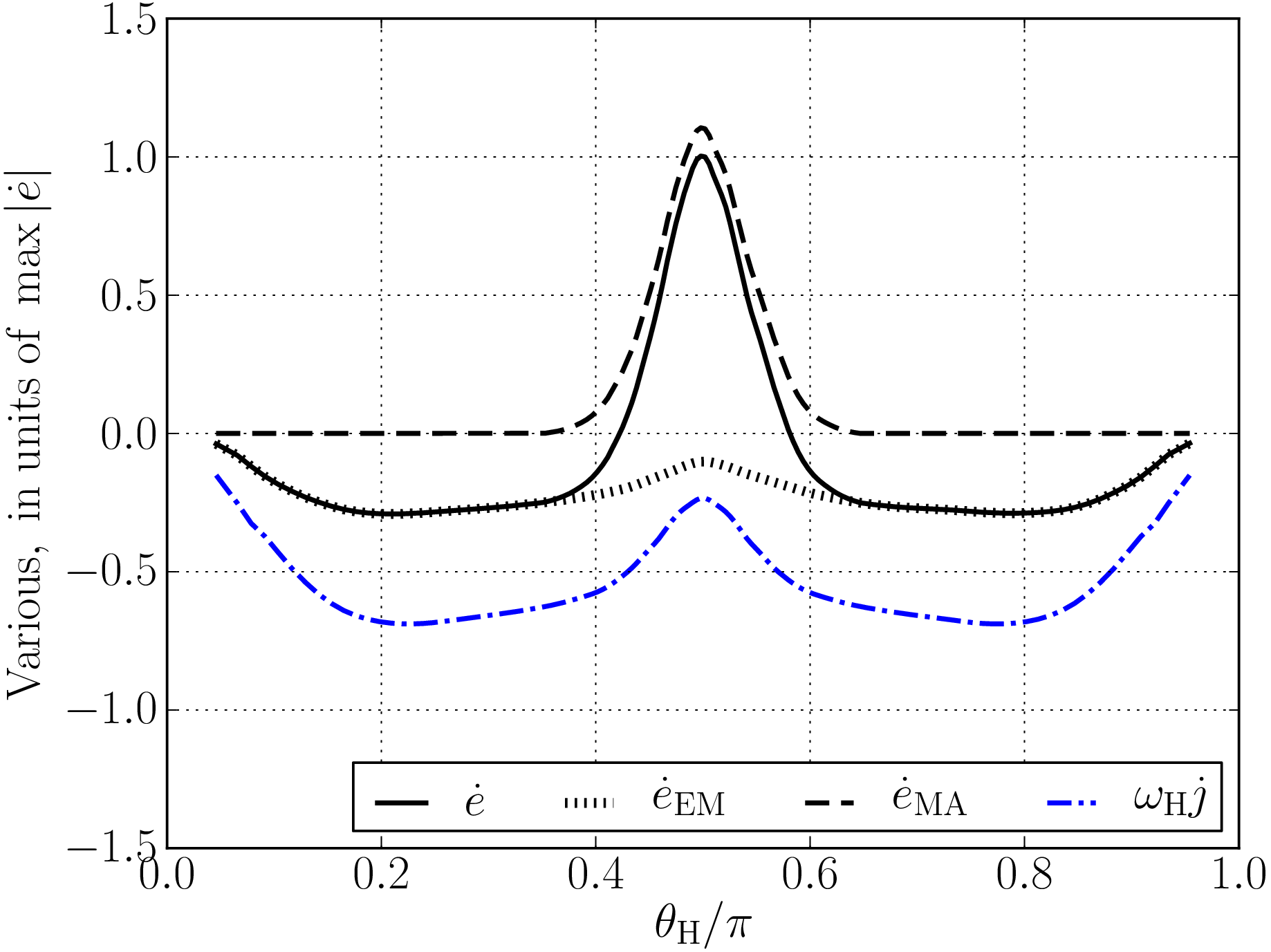}}
\caption{\small Same as in Fig. \ref{f:eminusomegal} for a MAD configuration. The black-hole spin is $a_*=0.99$. The electromagnetic energy density flux is everywhere negative on $\Sps_t$. The total energy density $\dot e $ is negative everywhere except in the equatorial belt where matter accretion dominates the energy balance. The condition $\dot \jmath < 0$ is satisfied everywhere on $\Sps_t$ (see the text for details).
}
\label{f:eminusomegal_mad}
\end{figure} 
In the GRMHD MAD simulations accretion of matter plays an essential role in accumulating magnetic field lines on the black hole, and contrary to the force-free case the energy-momentum of matter is not negligible. In Fig. \ref{f:eminusomegal_mad} in addition to the electromagnetic and matter energy density fluxes we plot the sum of the two representing the total energy flux. One can see that $\dot e$ is negative on the black-hole surface $\Sps_t$ except near the equator where energy absorption is dominated by matter accretion. Therefore the simulations of rotational energy extraction from a $a_*=0.99$ spinning black hole by a MAD field configuration satisfy the condition (\ref{e:Penrose_process}) on part of the black-hole surface and therefore describe a Penrose process involving electromagnetic fields. This is confirmed by the angular-momentum density flux being negative on the whole of the black hole surface. 
We see that the angular-momentum flux is negative over the entire horizon,  while the energy flux is negative only over the part of the surface exterior to the equatorial accretion flow. This is a characteristic property of the MAD configuration
because the rest-mass energy flux due to the accreted mass overwhelms the energy flux into the black hole and makes it positive, while this matter carries in very little angular momentum because its angular momentum is sub-Keplerian due to the action of strong magnetic fields that extract its angular-momentum and carry it away in the form of magnetized winds. 

To get more insight into the workings of the simulated black-hole rotational energy extraction process one has to leave the horizon and see what is happening in the bulk above the black-hole surface.

We have shown that GRMHD MAD simulations of black-hole rotational energy extraction describe a Penrose process but because of the approximations made we have not learned how this process works in detail. In the case of free particles we know what is happening: a particle decays in the ergoregion into one with negative and  another one with positive energies. The one with negative energy cannot leave the ergoregion and must be created there because negative energies exist only in the ergoregion  and energy along the trajectories is conserved. This cannot be the case for a perfect fluid (with nonzero pressure) or an electromagnetic field. However, the mechanical case can serve as a guide to what is happening in a more general case. For MAD simulations, one cannot expect to see negative energies in the ``bulk" since by stationarity energy is constant. However, the workings of the Penrose process should be apparent through the behavior of the Noether current $\vw P$.
Far from the black hole, the Noether current $\vw{P}$ is future directed timelike or null and is such that positive energy flows outwards. 
Near the black hole, in the ergoregion, $\vw{P}$ should become spacelike or past directed. This is indeed what is happening in our simulations.

\begin{figure}
\center
\centerline{\includegraphics[width=1.0\columnwidth]{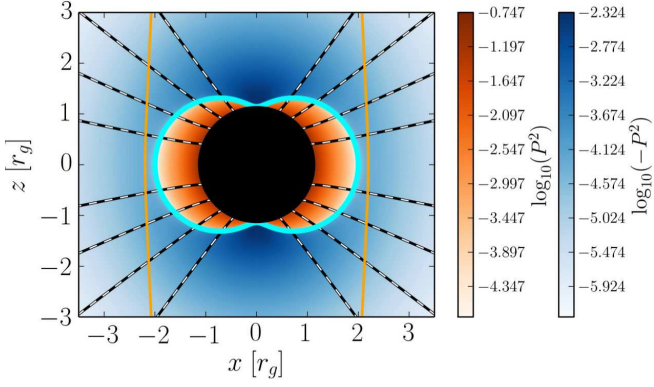}\hfill }
\center
\caption{\label{f:psqf} \small Color maps of $P^2$ in monopolar force-free
  spin $a_*=0.99$. The surface of the
  ergosphere is shown with cyan lines, the stagnation surface with
  orange lines. The region in which $\vw P$ is spacelike is shown in
  orange, and the region in which $\vw P$ is timelike is shown in blue
  (see color bar). Black-and-white striped lines represent the magnetic field lines. 
  As discussed in the main text, in a force-free configuration $\vw P$
  becomes null at the surface of the ergosphere.}
\end{figure}

\begin{figure}
\centerline{
\includegraphics[width=1.0\columnwidth]{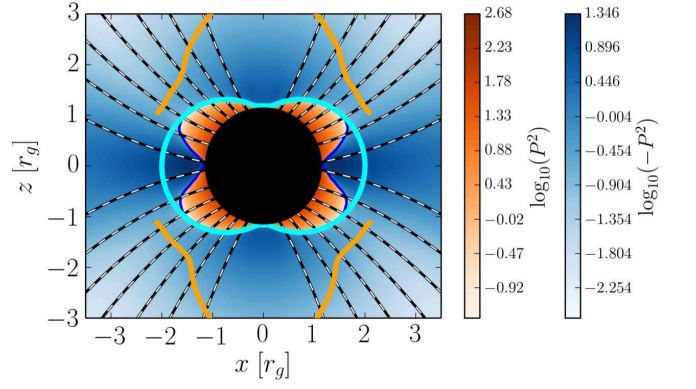}}
\center
\caption{\label{f:psqm} \small Color maps of $P^2$ in the MAD simulations for a black hole with
  spin $a_*=0.99$. Color codes and lines as in Fig. \ref{f:psqf}. In this case the surface
  $P^2=0$ nearly coincides with the surface of the ergosphere in
  magnetically-dominated polar jets, but lies inside of the surface of
  the ergosphere otherwise. }
\end{figure}

Figs.~\ref{f:psqf} and \ref{f:psqm} show the behavior of $\vw P$ in numerical results for the force-free and the MAD cases respectively.  We see that for a force-free configuration $P^2=0$ at the surface of the ergosphere whereas in the MAD
simulations the $P^2=0$ surface is very close to the surface of
the ergosphere in the polar jet regions but lies inside of it
elsewhere. These patterns are in full agreement with Figs. \ref{f:eminusomegal} and \ref{f:eminusomegal_mad}. They demonstrate the fundamental role played by the  ergoregion in extracting black-hole energy of rotation. This can be explained analytically as follows.

In the relativistic MHD code HARM, it is assumed that the Lorentz force on a
charged particle vanishes in the fluid frame:
\be
\label{eq:lorentz}
u_{\mu}F^{\mu\nu}=0.
\ee
Then a magnetic four-vector $b^{\mu}$ is defined as
\be
\label{eq:b}
b^{\mu}:= \frac{1}{2}\epsilon^{\mu\nu\alpha\beta}u_{\nu}F_{\alpha\beta},
\ee
with 
\be
\label{eq:bu}
b_{\mu}u^{\mu}=0,
\ee
following from ${\w F}$ antisymmetry.
 This allows the electromagnetic energy-momentum tensor (\ref{e:T_EM}) to be written in the form of \citep{Gammie03}:
\be
\label{eq:tmunumhd}
T^{\rm (EM)}_{\mu\nu} = b^2 u_\mu u_\nu + \frac{1}{2}b^2 g_{\mu\nu} - b_\mu b_\nu.
\ee

Therefore for the electromagnetic Noether current $P_\mu^{\rm (EM)}=T^{\rm (EM)}_{\mu\nu}\eta^{\nu}$ one has
\be
\label{eq:pff}
P^\mu_{(\rm EM)} P_\mu^{(\rm EM)} = P^2_{(\rm EM)} = \frac{1}{4} b^4 g_{tt}.  
\ee
Since $g_{tt} > 0$ inside ergosphere and $< 0$ outside, this fully
explains the numerical results seen in Fig.~\ref{f:psqf}:
\begin{eqnarray}
P^2_{(\rm EM)} &>& 0 \quad {\rm inside\ ergosphere},\\
P^2_{(\rm EM)} &<& 0 \quad {\rm outside\ ergosphere.}
\end{eqnarray}
Notice that this result applies not only to stationary axisymmetric
electromagnetic force-free field but also to time-dependent fully 3D
(nonaxisymmetric) configurations. However, the above property of $\vw P$ 
applies only to the electromagnetic force-free case. 

To see this let us use the general energy-momentum tensor
\[
T_{\mu\nu}= T^{\rm( MA)}_{\mu\nu} + T^{\rm (EM)}_{\mu\nu}.
\]
with $\w T^{\rm( MA)}$ and $\w T^{\rm (EM)}$ given by (\ref {eq:Tpf}) and (\ref{eq:tmunumhd})  respectively.
One obtains then
\be
\label{eq:p2general}
P^2 = \left(\frac{1}{2} b^2+p\right)^2 g_{tt} - A, 
\ee
with
\be A = 2(\Gamma-1) u b_t^2 + u_t^2 (\rho+u+p+b^2) [(2-\Gamma)u+\rho],
\ee
where $u=\epsilon - \rho$ is the internal energy and the adiabatic index $\Gamma$ ($p=(\Gamma - 1 )u$) satisfies $1 \le \Gamma \le 2$ (in the MAD simulations $\Gamma=4/3$).
For dust ($p=0$) one gets
\[
P^2= - \left(\rho u_t\right)^2,
\]
i.e., the Noether current is always timelike (but past directed for negative energy wordlines; see Sec. \ref{sub:mechP}).

For the force-free case ($b^2 \gg \rho$, $p\ll \rho$) one recovers (\ref{eq:pff}) but in general (e.g. for $\Gamma=4/3$) $A > 0$.

Since 
$P_{\rm EM}^2 = 0$ precisely at the surface of the ergosphere the same
applies to the full Noether current in the highly magnetized regions:
there $P^2 \approx P_{\rm EM}^2 = 0$ approximately at the ergosphere.  In the weakly
magnetized disk-corona region, however, $P^2=0$ will deviate from the
ergosphere by at least order unity. 
The first term on the right-hand side of Eq. (\ref{eq:p2general}) is positive
inside the ergosphere.  Since the second term is nonpositive for $1
\le\Gamma\le 2$ the surface $P^2 = 0$ lies \emph{inside} the
ergosphere as seen in Fig.~\ref{f:psqm}.

Also shown in Figs. ~\ref{f:psqf} and ~\ref{f:psqm} is the stagnation limit at which the field drift velocity
changes sign ($u^r=0$; inside this limit the velocity is pointing
inwards).  Inside the stagnation surface, an energy counterflow 
\cite{Komissarov-2009} is present: while the fields drift inwards, the
energy flows outwards. The stagnation limit is always outside the ergoregion; for $a_*=0.99$ it is very close to the ergosphere but for, e.g., $a_*=0.9999$ the two surfaces are still well separated. The shapes and location of our stagnation limits are different from those found by \citet{Okamoto-2006} and \citet{Komissarov-2009}. The reasons for these differences will be addressed in a future paper.

\section{Discussion and conclusions}
\label{section-Conclusions}

We proved that for any type of matter or (nongravitational) fields satisfying the weak energy condition, the black hole's rotational energy can be extracted if, and only if,
negative energy and angular momentum are absorbed by the black hole.  Applied to the case of a single particle, the general criterion (\ref{e:Penrose_process}) leads to the standard condition for a mechanical Penrose process. {For a general electromagnetic field, 
the criterion (\ref{e:Penrose_process}) leads to the condition (\ref{e:nec_cond_EM}) on the electromagnetic field at the horizon, which does not seem to have been expressed before.}

In a sense our findings are obvious (which does not mean they are trivial). They follow from the fact that the black-hole surface is a stationary null hypersurface. Hence it can only absorb matter or fields; it cannot emit anything, cannot emit energy. No torque can be applied to the horizon, since a torque results from a difference of material/field fluxes coming from the opposite sides of a surface \citep{Abramowicz-2010}. The only way to lose energy, independent of the nature of the medium the hole is interacting with,  is by absorbing a negative value of it. And, since the energy in question must be rotational, it must absorb negative angular momentum to slow it down.

Our results do not specify how the effect of net negative energy absorption by a black hole is achieved. The conditions for black-hole energy extraction do not guarantee the existence of such a process in the real world. As is well known, the mechanical Penrose process requires splitting of particles in the ergoregion but no realistic way of achieving black-hole energy extraction has been found. Using fluids (perfect or not) does not seem very promising in this context. The only known black-hole energy extracting process that might be at work in the Universe is the BZ mechanism. We showed that the process of energy extraction described by GRMHD simulations of magnetically arrested disk flows around rapidly spinning black holes is a Penrose process. This has been deduced before from energy conservation and efficiencies well in excess of 100\% but we showed that the solutions found by these simulations satisfy the rigorous and general conditions required by general relativity. Considering that black holes are purely general-relativistic objects this is a reassuring conclusion. 

It is worth stressing that when in the GRMHD simulations the Noether current has a positive flux in the outward direction everywhere (including at the BH horizon), it does not correspond to the flow of any {\sl physical} energy out of the black hole, since the ``energy" associated with the Noether current is not a measurable quantity: no physical observer can measure it, except at infinity, where the Killing vector $\eta$ becomes a unit timelike vector and therefore is eligible as the 4-velocity of a physical observer: an inertial observer at rest with respect to the BH location. 

As mentioned above, the main (and only important) difference between the mechanical and other versions of the Penrose process is that in the first version, particles move along geodesics and therefore energy is conserved on their trajectories. Therefore the motion of a particle crossing the horizon with negative energy is from its start restricted to the ergoregion. This does not have to be the case of interacting matter and fields. It is still true that the ``outgoing flow of energy at infinity in the Penrose process is inseparable from the negative energy at infinity of an infalling `object' "\citep[to quote][]{Komissarov-2009}, but this inseparability concerns the negative energy of the object when it is absorbed by the black hole. On its way to the final jump into the hole, the object's energy may vary depending on its interactions with the medium it is part of.

A detailed description of these processes in the framework of the GRMHD simulations will be the subject of a future work.

\begin{acknowledgments}
MA and JPL thank Serguei Komissarov for an enlightening and stimulating exchange of Emails.
Research reported here was partially supported by Polish NCN Grants No. UMO-2011/01/B/ST9/05439, No. UMO-2011/01/B/ST9/05437, and 
No. DEC-2012/04/A/ST9/00083. 
Research at the Silesian University in Opava was supported by
the Czech CZ.1.07/2.3.00/20.0071 ``Synergy'' grant for international collaboration. JPL acknowledges a grant from the French Space Agency CNES 
and EG, Grant No. ANR-12-BS01-012-01 ``Analyse Asymptotique en Relativit\'e G\'en\'erale''
from Agence Nationale de la Recherche. 
Support for this work was provided by a Princeton Center for Theoretical Science
Fellowship and by NASA through Einstein Postdoctoral Fellowship Grant No. PF3-140115 awarded by the Chandra X-ray Center, which is operated by the Smithsonian Astrophysical Observatory for NASA under Contract No. NAS8-03060.
We acknowledge support by the NSF through TeraGrid/XSEDE resources provided by NICS Kraken and LONI QueenBee, where simulations were carried out; NICS Nautilus, where data were analyzed; and TACC Ranch and NCSA MSS, where data were backed up, under Grants No. TG-AST100040 (A.T.) and TG-AST080026N (R.N.).
\end{acknowledgments}

\appendix
\setcounter{section}{0}
\setcounter{footnote}{0}

\section{Kerr solution in 3+1 Kerr coordinates}
\label{ap:Kerr}
The Kerr solution is described by two parameters: the mass $m$ and the specific
angular momentum $a:=J/m$, $J$ being the total angular momentum. 
The metric components with with respect to the ``3+1''
Kerr coordinates $(t,r,\theta,\varphi)$ are given by (see e.g. \cite{Gourgoulhon-2005})
\bea
  g_{\mu\nu} \D x^\mu \D x^\nu  =   - \left( 1 - \frac{2mr}{\rho^2} \right) \D t^2 + \frac{4mr}{\rho^2}\, \D t\, \D r && \nonumber \\
    - \frac{4amr}{\rho^2} \sin^2\theta \, \D t\, \D \varphi +  \left( 1 + \frac{2mr}{\rho^2} \right)  \D r^2 && \nonumber \\
    - 2 a \sin^2\theta \left( 1+ \frac{2mr}{\rho^2} \right) \D r\, \D \varphi + \rho^2 \D \theta^2 &&
   \nonumber \\
     + \left( r^2 + a^2 + 
  \frac{2 a^2 m r\sin^2\theta}{\rho^2}\right)\sin^2\theta
   \, \D \varphi^2 ,   &&   \label{e:KE:metric_comp_Kerr}
\eea
with 
\be
    \rho^2 := r^2 + a^2 \cos^2\theta .  \label{e:KE:rho_def}
\ee
The coordinates $(t,r,\theta,\varphi)$ are a 3+1 version 
of the original Kerr coordinates \citep{Kerr-1963} and can be viewed as
a spheroidal version
of the well-known ``Cartesian" Kerr-Schild coordinates.
The event horizon $\Hor$ is located at 
\be 
    r = r_H := m + \sqrt{m^2-a^2} . \label{e:KE:r_H_def}
\ee
and the black-hole angular velocity $\omega_H$ defined by (\ref{e:ell_eta_xi})
takes the value $\omega_H=a/(2mr_H)$.
Since $r_H$ does not depend upon $\theta$ nor $\varphi$, 
the Kerr coordinates are adapted to $\Hor$, in the sense
defined in Sec.~\ref{s:adapted_coord}. 

Note that the metric components given by Eq.~(\ref{e:KE:metric_comp_Kerr})
are all regular at $r=r_H$.\footnote{On the contrary, most of them are
singular at $\rho=0$, which, via (\ref{e:KE:rho_def}),
corresponds to $r=0$ and $\theta=\pi/2$. In   
Kerr-Schild coordinates, this corresponds to the ring $x^2+y^2=a^2$ in the
plane $z=0$. This is the ring singularity of Kerr spacetime.}
Note also that in the limit $a\rightarrow 0$, then $\rho\rightarrow r$ and
the line element (\ref{e:KE:metric_comp_Kerr})  reduces
to the Schwarzschild metric in 3+1 Eddington-Finkelstein coordinates. 

From (\ref{e:KE:metric_comp_Kerr}), one can compute the determinant $g$ of
the metric with respect to the 3+1 Kerr coordinates and get the relatively 
simple expression:
\be \label{e:det_g_Kerr}
    \sqrt{-g} = (r^2 + a^2\cos^2\th) \sin\th . 
\ee

The metric (\ref{e:KE:metric_comp_Kerr}) is clearly 
stationary and axisymmetric and the two vectors
\be \label{e:KE:Killing_vect}
    \vw{\eta} := \left( \der{}{t} \right) _{r,\theta,\varphi}
    \qquad \mbox{and} \qquad
    \vw{\xi}  := \left( \der{}{\varphi} \right) _{t,r,\theta}
\ee
are the two Killing vectors, $\vw{\eta}$ being associated with the
stationarity and $\vw{\xi}$ with the axial symmetry of the Kerr
spacetime. These two Killing vectors are identical to
the ``standard'' Killing vectors which are formed using the
Boyer-Lindquist coordinates $(t_{\rm BL},r,\theta,\varphi_{\rm BL})$:
\be
    \vw{\eta}= \left( \der{}{t_{\rm BL}} \right) _{r,\theta,\varphi_{\rm BL}}
    \qquad \mbox{and} \qquad
    \vw{\xi}= \left( \der{}{\varphi_{\rm BL}} \right) _{t_{\rm BL},r,\theta} .
\ee

The Killing vector $\vw{\eta}$ ceases to be timelike at the boundary of the ergoregion
(the ergosphere),
\be
\label{e:ergos}
r_{\rm erg} = m + \sqrt{m^2 - a^2\cos^2 \theta}, 
\ee
below which it is spacelike ($g_{tt}=\vw{\eta}\cdot\vw{\eta} >0$).

The angular speed of the dragging of inertial frames can be written as
\be
\label{e:omega_drag}
\omega=\frac{\vw{\eta}\cdot \vw{\xi}}{\vw{\xi}\cdot \vw{\xi}}=\frac{g_{t\phi}}{g_{\phi\phi}}=\frac{2Jr}{A}=\frac{2amr}{A}
\ee
where $A=(r^2 + a^2) - \Delta\,a^2\sin^2\theta$ with $\Delta=r^2 - 2mr + a^2$.
At the horizon $\Delta=0$ and $\omega=\omega_H$.

Setting $\D r =0$ and $r = r_H$ in the line element (\ref{e:KE:metric_comp_Kerr})
yields the metric $\w{\gamma}_H$ induced on $\Hor$:
\begin{widetext}
\be \label{e:metric_H}
     (\gamma_H)_{AB} \D x^A \D x^B = 2mr_H 
     \left[ 
        (1-a\omega_H\sin^2\th) \, \D\th^2  + 
        \frac{\sin^2\th}{1-a\omega_H\sin^2\th} \, (\D\phi - \omega_H\D t)^2 \right] ,
\ee
\end{widetext}
where $(x^A)$ stands for the coordinates spanning $\Hor$:
$(x^A) = (t,\th,\phi)$. This metric is clearly degenerate, with the 
degeneracy direction along $\ell^A = (1,0,\omega_H)$. We thus recover the fact
that $\Hor$ is a null hypersurface. 

Setting $\D t=0$ in the line element (\ref{e:metric_H}), we get the 
induced metric $\w{q}$ in the 2-surfaces $\Sps_t$ that foliate $\Hor$: 
\begin{widetext}
\be \label{e:metric_St}
q_{ab} \D x^a \D x^b = 2mr_H 
  \left[ 
        (1-a\omega_H\sin^2\th) \, \D\th^2 + 
       \frac{\sin^2\th}{1-a\omega_H\sin^2\th} \, \D\phi^2 \right],
\ee
\end{widetext}
where $(x^a)$ stands for the coordinates spanning $\Sps_t$:
$(x^a) = (\th,\phi)$. 
The metric $\w{q}$ is clearly positive definite, hence the 2-surfaces are
spacelike. From (\ref{e:metric_St}), we read immediately 
the determinant of $\w{q}$ with respect to the coordinates $(\th,\phi)$: 
\be \label{e:sqrt_det_q}
    \sqrt{q} = 2 m r_H \sin\th . 
\ee

\section{Flux integrals on a hypersurface}
\label{ap:flux_integrals}

Let $\Sigma$ be an oriented hypersurface in the spacetime $(\M,\w{g})$. 
From the very definition of the integral of a 3-form over a three-dimensional 
manifold, we have
\bea \label{e:flux_parallelepiped}
    \int_\Sigma \weps(\vw{P})& =& 
    \int_\Sigma \weps(\vw{P})(\D\vw{x}_{(1)}, \D\vw{x}_{(2)}, \D\vw{x}_{(3)})\nonumber \\
   & &= \int_\Sigma \weps(\vw{P}, \D\vw{x}_{(1)}, \D\vw{x}_{(2)}, \D\vw{x}_{(3)}) , 
\eea
where the last equality follows from the definition (\ref{e:def_epsP})
of $\weps(\vw{P})$ and $(\D\vw{x}_{(1)}, \D\vw{x}_{(2)}, \D\vw{x}_{(3)})$
are infinitesimal vectors forming an elementary
right-handed parallelepiped on $\Sigma$. 

\subsection{Case of a spacelike or timelike hypersurface}

If $\Sigma$ is spacelike or timelike, we may introduce the unit normal
$\vw{m}$ that is compatible with $\Sigma$'s orientation (i.e. such that
the orientation is given by the 3-form $\weps(\vw{m}) = \weps(\vw{m},.,.,.)$,
cf. Sec.~\ref{s:energy_conservation}). 
The orthogonal decomposition of $\vw{P}$ with respect to $\Sigma$ is then
\be \label{e:P_ortho_decomp}
    \vw{P} = \pm (P_\mu m^\mu) \,  \vw{m} + \vw{P}_{||} ,
\ee
where $\pm$ is $+$ (resp. $-$) if $\Sigma$ is timelike 
(resp. spacelike) and $\vw{P}_{||}$ is tangent to $\Sigma$. 
The four vectors $\vw{P}_{||}$, $\D\vw{x}_{(1)}$, $\D\vw{x}_{(2)}$ 
and $\D\vw{x}_{(3)}$ cannot be linearly independent, being all tangent to $\Sigma$,
so that 
$\weps(\vw{P}_{||}, \D\vw{x}_{(1)}, \D\vw{x}_{(2)}, \D\vw{x}_{(3)})=0$. 
Hence 
\bea \label{e:espP_parallelepiped}
\weps(\vw{P}, \D\vw{x}_{(1)}, \D\vw{x}_{(2)}, \D\vw{x}_{(3)})
= \qquad   \qquad   \qquad  \qquad   &&  \nonumber \\
\qquad   \qquad    \pm (P_\mu m^\mu) \, 
 \weps(\vw{m}, \D\vw{x}_{(1)}, \D\vw{x}_{(2)}, \D\vw{x}_{(3)})&&
\eea
Now, since $\vw{m}$ is a unit vector,  
\be \label{e:dV_parallelepiped}
    \D V := \weps(\vw{m}, \D\vw{x}_{(1)}, \D\vw{x}_{(2)}, \D\vw{x}_{(3)})
\ee
is nothing but the volume of the elementary parallelepiped
formed by $(\D\vw{x}_{(1)}, \D\vw{x}_{(2)}, \D\vw{x}_{(3)})$ with respect to
the 3-metric $\w{\gamma}$ induced by $\w{g}$ on $\Sigma$
(for $\Sigma_{\rm ext}$, $\w{\gamma}$ is denoted by $\w{h}$ in 
(\ref{e:E_ext})). 
By combining (\ref{e:flux_parallelepiped}), (\ref{e:espP_parallelepiped})
and (\ref{e:dV_parallelepiped}), we get 
\be
    \int_\Sigma \weps(\vw{P}) = \pm \int_{\Sigma}  P_\mu m^\mu \, \D V . 
\ee
This establishes the second equalities in 
(\ref{e:def_E1}), (\ref{e:def_E2}) and (\ref{e:E_ext}). 

Let $(x^1,x^2,x^3)$ be a coordinate system on $\Sigma$ and let us choose 
the $\D\vw{x}_{(i)}$'s as the corresponding elementary displacements:
\[
    \D\vw{x}_{(1)} =  \D x^1 \der{}{x^1}, \quad
   \D\vw{x}_{(2)} =  \D x^2 \der{}{x^2}, \quad 
   \D\vw{x}_{(3)} =  \D x^3 \der{}{x^3} . 
\]
Then 
\be
    \D V = \sqrt{|\gamma|} \, \D x^1 \, \D x^2 \, \D x^3 ,  
\ee
where $\gamma = \det(\gamma_{ij})$, the $\gamma_{ij}$'s being the components
of the induced 3-metric on $\Sigma$. 
This established the third equalities in (\ref{e:def_E1}), (\ref{e:def_E2}) and (\ref{e:E_ext}). 

\subsection{Case of null hypersurface}
\label{ap:flux_integrals_null}

Here we consider that $\Sigma = \Delta\Hor$, but the results are valid for
any null hypersurface. Since $\Delta\Hor$ is null, there is no orthogonal 
decomposition of $\vw{P}$ of the type (\ref{e:P_ortho_decomp}). 
Let us consider instead the slicing of $\Delta\Hor$ by the 2-spheres $\Sps_t$
of constant $t$ (cf. Sec.~\ref{sub-section-Kerr-horizon}). 
Then we have the following unique decomposition of $\vw{P}$: 
\be \label{e:P_null_decomp}
    \vw{P} = - (P_\mu \ell^\mu) \, \vw{k} - (P_\mu k^\mu) \, \vw{\ell} + \vw{P}_{||} , 
\ee
with $\vw{P}_{||}$ is tangent to $\Sps_t$. This decomposition follows from the fact 
that $\vw{k}$ and $\vw{\ell}$ generate the 2-plane orthogonal to $\Sps_t$ and from the normalization relation (\ref{e:k_ell}). 

Let us choose the elementary parallelepiped $(\D\vw{x}_{(1)}, \D\vw{x}_{(2)}, \D\vw{x}_{(3)})$ on $\Delta\Hor$ such that 
\[
     \D\vw{x}_{(1)}  = \D t \, \vw{\ell}   
\]
and $\D\vw{x}_{(2)}$ and $\D\vw{x}_{(3)}$ are tangent to $\Sps_t$. 
The integrand in (\ref{e:flux_parallelepiped}) is then 
\be
    \weps(\vw{P}, \D\vw{x}_{(1)}, \D\vw{x}_{(2)}, \D\vw{x}_{(3)}) 
   =  \D t \, \weps(\vw{P}, \vw{\ell}, \D\vw{x}_{(2)}, \D\vw{x}_{(3)})
\ee
Now, from (\ref{e:P_null_decomp}), 
\bea
    \weps(\vw{P},\vw{\ell},\D\vw{x}_{(2)}, \D\vw{x}_{(3)})& = &
  - (P_\mu \ell^\mu) \, \weps(\vw{k},\vw{\ell},\D\vw{x}_{(2)}, \D\vw{x}_{(3)}) \nonumber \\
  && - (P_\mu k^\mu) \underbrace{\weps(\vw{\ell},\vw{\ell},\D\vw{x}_{(2)}, \D\vw{x}_{(3)})}_{0}\nonumber \\
  &&+ \underbrace{\weps(\vw{P}_{||},\vw{\ell},\D\vw{x}_{(2)}, \D\vw{x}_{(3)})}_{0} \nonumber  \\
  && = - (P_\mu \ell^\mu) \, \weps(\vw{k},\vw{\ell},\D\vw{x}_{(2)}, \D\vw{x}_{(3)}).  \nonumber
\eea
Therefore we may rewrite (\ref{e:flux_parallelepiped}) as
\be
    \int_{\Delta\Hor} \weps(\vw{P}) = - \int_{\Delta\Hor} P_\mu \ell^\mu
    \, \D V , 
\ee
with 
\be \label{e:dV_null}
    \D V = \weps(\vw{k},\vw{\ell},\D\vw{x}_{(2)}, \D\vw{x}_{(3)}) \, \D t . 
\ee
This establishes the second equality in (\ref{e:E_H}). 

Let $(y^1,y^2)$ be a coordinate system on $\Sps_t$ and let us choose 
the $\D\vw{x}_{(2)}$ and $\D\vw{x}_{(3)}$ as the corresponding elementary 
displacements:
\[
    \D\vw{x}_{(2)} =  \D y^1 \der{}{y^1}, \quad
   \D\vw{x}_{(3)} =  \D y^2 \der{}{y^2}, \quad 
\]
We have then 
\bea \label{e:dV_null_prov}
   \D V &=&  \weps(\vw{k},\vw{\ell},\D\vw{x}_{(2)}, \D\vw{x}_{(3)}) \, \D t \nonumber \\
   & =& \weps(\vw{k},\vw{\ell},\dert{}{y_1},\dert{}{y_2}) 
    \; \D t \, \D y^1 \, \D y^2  \nonumber \\
    &=& \sqrt{-{\tilde g}} \, \D t\, \D y^1 \, \D y^2 , 
\eea 
where ${\tilde g}$ is the determinant of the components of $\w{g}$ in the basis $(\vw{k},\vw{\ell},\dert{}{y_1},\dert{}{y_2})$. Given the definitions of $\vw{k}$ and $\w{q}$, 
these components are 
\be \label{e:gab_null}
  {\tilde g}_{\alpha\beta} = \left(\begin{array}{cccc}
  0 & -1 & 0 & 0 \\
  -1 & 0 & 0 & 0  \\
  0 & 0 & q_{11} & q_{12} \\ 
  0 & 0 & q_{12} & q_{22} 
  \end{array} \right) . 
\ee
Hence $\tilde g = - q$, with $q := \det(q_{ab})$, and 
(\ref{e:dV_null_prov}) becomes
\be
    \D V = \sqrt{q}  \, \D t\, \D y^1 \, \D y^2 . 
\ee
This establishes the third equality in (\ref{e:E_H}). 

\section{Calculation of particle energy as a flux through some hypersurface}
\subsection{Case of a spacelike hypersurface}
\label{ap:Part_sp}

As shown in Sec.~\ref{sub:mechP}, the particle energy at the event $A_1$ on $\Sigma_1$ is
\begin{widetext}
\begin{equation}
    E_1 = \mathfrak{m}_1 \int_{\Sigma_1} \int_{-\infty}^{+\infty} 
        \delta_{A(\tau)}(M)\; 
        g_\mu^{\ \, \rho}(M,A(\tau)) (u_1)_\rho(\tau) \;  
   g_\nu^{\ \, \sigma}(M,A(\tau))  (u_1)_\sigma(\tau)  \;  
                     \eta^\mu(M) \, n_1^\nu(M)   \sqrt{\gamma}
        \, \D x^1 \, \D x^2 \, \D x^3\, \D\tau .
\end{equation}
\end{widetext}
Thanks to the Dirac distribution, only the terms for which $M=A(\tau)$ contribute to the above integral. 
We may then drop the parallel propagators and write 
\begin{widetext}
\[
    E_1 = \mathfrak{m}_1 \int_{\Sigma_1} \int_{-\infty}^{+\infty} 
        \delta_{A(\tau)}(M)\; (u_1)_\mu(\tau) \, \eta^\mu(M) \;
        (u_1)_\nu(\tau) \, n_1^\nu(M) \sqrt{\gamma}
        \, \D x^1 \, \D x^2 \, \D x^3\, \D\tau .
\]
\end{widetext}
Let us introduce in the vicinity of $A_1$ a coordinate system $(t,x^1,x^2,x^3)$ such that
$\Sigma_1$ is the hypersurface $t=0$ and $t$ increases towards the future. Then the normal $\vw{n}_1$ is collinear to the gradient of $t$:
$(n_1)_\alpha = - N \nabla_\alpha t$, the coefficient $N>0$ being called the \emph{lapse function}. 
We have then $(n_1)_\alpha = (-N,0,0,0)$ and
\[
   (u_1)_\nu\, n_1^\nu = (n_1)_\nu u_1^\nu = -N u_1^0 . 
\]
Hence
\bea
    E_1 &=& - \mathfrak{m}_1 \int_{\Sigma_1} \int_{-\infty}^{+\infty} 
        \delta_{A(\tau)}(M)\; \eta_\mu(M) \, u_1^\mu(\tau)  \,\nonumber \\
      && \qquad \qquad \qquad \qquad\times N \sqrt{\gamma}
        \, \D x^1 \, \D x^2 \, \D x^3\, u_1^0 \D\tau . \nonumber
\eea
Since the particle's worldline is timelike and therefore never tangent to $\Sigma_1$, we may use $t$ as a regular parameter along it and 
perform the change of variable $\tau \rightarrow t$ in the above integral. Taking into account that
$u_1^0 = \D t / \D\tau$ (from the very definition of a 4-velocity), we get 
\bea
    E_1 &=& - \mathfrak{m}_1 \int_{\Sigma_1} \int_{-\infty}^{+\infty} 
        \delta_{A(t)}(M)\; \eta_\mu(M) \, u_1^\mu(t) \, \nonumber \\
  && \qquad \qquad \qquad \qquad\times N \sqrt{\gamma}
        \, \D x^1 \, \D x^2 \, \D x^3\, \D t .
\eea
Within the coordinate system $(t,x^1,x^2,x^3)$, the coordinates of $A(t)$ are $(t,z^1(t),z^2(t),z^3(t))$
and those of $M$ are $(0,x^1,x^2,x^3)$ (for $M\in\Sigma_1$). 
Therefore using (\ref{e:delta_A_M_explicit}) along with the identity $\sqrt{-g} = N \sqrt{\gamma}$
(see e.g. Eq.~(5.55) in \cite{Gourg12}), we obtain
\bea
    E_1 = - m_1 \int_{\Sigma_1} \int_{-\infty}^{+\infty} 
        \delta(-t) \,\delta(x^1-z^1(t)) \,\delta(x^2-z^2(t)) \, && \nonumber \; \\
         \times \delta(x^3-z^3(t))\, \eta_\mu(M) \, u_1^\mu(t)
        \, \D x^1 \, \D x^2 \, \D x^3\, \D t . &&  \nonumber
\eea
Since $\delta(-t)=\delta(t)$, performing the integration on $t$ leads to 
\begin{widetext}
\be
    E_1 =  - \mathfrak{m}_1 \int_{\Sigma_1} 
        \delta(x^1-z^1(0)) \,\delta(x^2-z^2(0)) \,\delta(x^3-z^3(0))  
         \, \eta_\mu(M) \, u_1^\mu(0)
        \, \D x^1 \, \D x^2 \, \D x^3 \nonumber 
         =  - \mathfrak{m}_1 \; \eta_\mu(0,z^1(0),z^2(0),z^3(0)) \; u_1^\mu(0) . \nonumber 
\ee
\end{widetext}
Since $(0,z^1(0),z^2(0),z^3(0))$ are the coordinates of $A_1$ and $u_1^\mu(0)$ are the components of 
$\vw{u}_1$ at $A_1$, we conclude that 
\be \label{e:E_1_particle_append}
    E_1 = - \mathfrak{m}_1  \left. (\eta_\mu u_1^\mu) \right| _{A_1} = - \mathfrak{m}_1 \, \eta_\mu u_1^\mu .
\ee

\subsection{Case of a null hypersurface}
\label{ap:Part_null}

In Sec.~\ref{sub:mechP} we obtained for the energy of particle crossing the event horizon
\begin{widetext}
\be \label{e:E_H_particle_prov}
 \Delta E_H = \mathfrak{m}_* \int_{\Delta\Hor} \int_{-\infty}^\infty \delta_{A(\tau)}(M) \,  (u_*)_\mu(\tau) \eta^\mu(M) \;  
(u_*)_\nu(\tau) \ell^\nu(M) \, \sqrt{q} \, \D t \, \D y^1 \, \D y^2 \, \D\tau . 
\ee
\end{widetext}
Note that, for the same reasons as above, we have dropped the parallel propagators.
Let us introduce in the vicinity of $A_H$ a coordinate system $(w,t,y^1,y^2)$ such that 
$\Hor$ is the hypersurface $w=0$, $\vw{k} = \dert{}{w}$ on $\Hor$ and $\vw{\ell} = \dert{}{t}$ on $\Hor$. 
Let us expand $\vw{u}_*$ in the associated coordinate basis:  
\[
    \vw{u}_* = u_*^0 \, \vw{k} + u_*^1 \, \vw{\ell} + u_*^2 \, \der{}{y^1} + u_*^3 \, \der{}{y^2} . 
\]
We have then, given (\ref{e:k_ell}) and the orthogonality of $\vw{\ell}$ to itself and to 
$\dert{}{y^1}$ and $\dert{}{y^2}$, 
\be \label{e:us_ell}
    (u_*)_\nu \ell^\nu = u_*^\nu \ell_\nu = - u_*^0 = - \frac{\D w}{\D\tau} . 
\ee
Since the worldline of $\Pp_*$ is crossing $\Hor$, we may use $w$ as a regular parameter on it and perform
the change of variable $\tau \rightarrow w$ in the integral (\ref{e:E_H_particle_prov}), taking 
advantage of (\ref{e:us_ell}). Therefore
\bea
   \Delta E_H = \qquad \qquad\qquad \qquad\qquad \qquad \qquad \qquad\qquad \qquad \quad &&\nonumber \\
   - \mathfrak{m}_* \int_{\Delta\Hor} \int_{-\infty}^{+\infty} \delta_{A(w)}(M) \,  \eta_\mu(M) u_*^\mu(w)  \;  
 \, \sqrt{q} \, \D t \, \D y^1 \, \D y^2 \, \D w . && \nonumber
\eea
Within the coordinate system $(w,t,y^1,y^2)$, the coordinates of $A(w)$ are $(w,z^1(w),z^2(w),z^3(w))$
and those of $M$ are $(0,t,y^1,y^2)$ (for $M\in\Delta\Hor$). 
Therefore, using (\ref{e:delta_A_M_explicit}), we obtain
\bea
        \Delta E_H & = & - \mathfrak{m}_* \int_{\Delta\Hor} \int_{-\infty}^{+\infty} 
        \delta(-w) \,\delta(t-z^1(w)) \,\delta(y^1-z^2(w)) \,  
          \nonumber  \\
         & & \times \, \delta(y^2-z^3(w)) \eta_\mu(M) \, u_*^\mu(w) \frac{\sqrt{q}}{\sqrt{-g}} \, \D t \, \D y^1 \, \D y^2 \, \D w .
            \nonumber 
\eea
Performing the integration on $w$, we get
\bea
    \Delta E_H &=& - \mathfrak{m}_* \int_{\Delta\Hor} 
        \,\delta(t-z^1(0)) \,\delta(y^1-z^2(0)) \,\delta(y^2-z^3(0))   \nonumber\\
     && \qquad \qquad \qquad \qquad\times    \eta_\mu(M) \, u_*^\mu(0) \,  
         \, \frac{\sqrt{q}}{\sqrt{-g}} \, \D t \, \D y^1 \, \D y^2 .  \nonumber
      \eea
On $\Delta\Hor$, the components of the metric tensor with respect to the 
coordinates $(w,t,y^1,y^2)$ are given by (\ref{e:gab_null}), from which we deduce that
$\sqrt{-g} = \sqrt{q}$. Noticing that $(0,z^1(0),z^2(0),z^3(0))$ are the coordinates of $A_H$, we
conclude that
\be 
    \Delta E_H = - \mathfrak{m}_* \left. (\eta_\mu u_*^\mu) \right| _{A_H} = - \mathfrak{m}_* \, \eta_\mu u_*^\mu . 
\ee

\section{Energy and angular-momentum conservation laws in adapted coordinates}
\label{ap:spherical}

In this appendix, we derive the energy conservation law (\ref{e:ener_cons}), 
as well as the angular-momentum one (\ref{e:angu_mom}), 
by a direct calculation within \emph{adapted} coordinates 
$(x^\alpha)=(t,r,\th,\phi)$,
as defined in Sec.~\ref{s:adapted_coord}. 
The starting point is the covariant energy-momentum conservation law 
$\nabla_\mu T^\mu_{\ \, \alpha} = 0$, which can be expressed in terms of partial
derivatives thanks to a standard formula for the covariant divergence of a symmetric tensor 
tensor field: 
\be
    \frac{1}{\sqrt{-g}} \der{}{x^\mu} \left( \sqrt{-g} T^\mu_{\ \, \alpha} \right)
        - \frac{1}{2} \der{g_{\mu\nu}}{x^\alpha} T^{\mu\nu} = 0 . 
\ee
For $\alpha=0$ and $\alpha=3$, the second term in the left-hand side vanishes, due
to the spacetime symmetries ($\dert{g_{\mu\nu}}{t}=0$ and $\dert{g_{\mu\nu}}{\phi}=0$). 
We are thus left with
\be
    \der{}{x^\mu} \left( \sqrt{-g} T^\mu_{\ \, \alpha} \right) = 0 
    \quad (\alpha=0,3).
\ee
Let us integrate this equation over the 
coordinate 4-volume formed by the Cartesian product
$[t_1,t_2]\times [r_H,r_{\rm ext}]\times[0,\pi]\times[0,2\pi]$. This 
corresponds to the coordinate ranges of the spacetime 4-volume
enclosed in the hypersurface 
$\VV := \Sigma_1 \cup \Delta\Hor \cup \Sigma_2 \cup \Sigma_{\rm ext}$
considered in Sec.~\ref{s:conservation_laws}
and to which the coordinates $(t,r,\th,\phi)$ are adapted. We get
\begin{widetext}
\be \label{e:integral_4D_coord}
    \int_{t=t_1}^{t=t_2} \int_{r=r_H}^{r=r_{\rm ext}} \int_{\th=0}^{\th=\pi}
    \int_{\phi=0}^{\phi=2\pi} 
    \der{}{x^\mu} \left( \sqrt{-g} T^\mu_{\ \, \alpha} \right) 
    \, \D t \,  \D r \, \D \th \, \D \phi
     = 0 \,  (\alpha=0,3).
\ee
Since the integral bounds are independent from one another, we may permute
the various integrals and use the identities
\be \label{e:integral_t}
    \int_{t=t_1}^{t=t_2} \der{}{t} \left( \sqrt{-g} T^t_{\ \, \alpha} \right)
        \D t =  \left( \sqrt{-g} T^t_{\ \, \alpha} \right)_{t=t_2}
            - \left( \sqrt{-g} T^t_{\ \, \alpha} \right)_{t=t_1} 
\ee
\be \label{e:integral_r}
    \int_{r=r_H}^{r=r_{\rm ext}} \der{}{r} \left( \sqrt{-g} T^r_{\ \, \alpha} \right)
        \D r =  \left( \sqrt{-g} T^r_{\ \, \alpha} \right)_{r=r_{\rm ext}}
            - \left( \sqrt{-g} T^r_{\ \, \alpha} \right)_{r=r_H} 
\ee
\be \label{e:integral_theta}
    \int_{\th=0}^{\th=\pi} \der{}{\th} \left( \sqrt{-g} T^\th_{\ \, \alpha} \right)
        \D \th =  \left( \sqrt{-g} T^\th_{\ \, \alpha} \right)_{\th=\pi}
            - \left( \sqrt{-g} T^\th_{\ \, \alpha} \right)_{\th=0} = 0 
\ee
\be \label{e:integral_phi}
    \int_{\phi=0}^{\phi=2\pi} \der{}{\phi} \left( \sqrt{-g} T^\phi_{\ \, \alpha} \right)
        \D \phi =  \left( \sqrt{-g} T^\phi_{\ \, \alpha} \right)_{\phi=2\pi}
            - \left( \sqrt{-g} T^\phi_{\ \, \alpha} \right)_{\phi=0} = 0 .  
\ee
The ``$=0$'' in (\ref{e:integral_theta}) results from $\sqrt{-g} = 0$ 
at $\th=0$ and $\th=\pi$, as a consequence of regularity properties of 
spherical coordinates, while the ``$=0$'' of (\ref{e:integral_phi}) 
results from the $2\pi$-periodicity associated with the coordinate $\phi$.
Taking into account (\ref{e:integral_t})-(\ref{e:integral_phi}), Eq. 
(\ref{e:integral_4D_coord}) becomes
\bea
  \int_{\Sigma_2}  T^t_{\ \, \alpha} \sqrt{-g} \,  \D r \, \D \th \, \D \phi
    \ - \ \int_{\Sigma_1}  T^t_{\ \, \alpha} \sqrt{-g} \, \D r \, \D \th \, \D \phi 
   + \ \int_{\Sigma_{\rm ext}} T^r_{\ \, \alpha} \sqrt{-g} \,  \D t \, \D \th \, \D \phi 
   \  - \int_{\Delta\Hor} T^r_{\ \, \alpha} \sqrt{-g} \,  \D t \, \D \th \, \D \phi 
    = 0   \nonumber \\ \quad (\alpha=0,3).
\eea
For $\alpha=0$, we recognize the energy conservation law (\ref{e:ener_cons}), 
the four integrals being, respectively, $-E_2$, $E_1$, $-\Delta E_{\rm ext}$
and $-\Delta E_H$ as given by (\ref{e:E_1_E_2_adapted})-(\ref{e:E_ext_adapted}). 
For $\alpha=3$, we get the angular-momentum conservation law (\ref{e:angu_mom}),
the four integral being respectively $J_2$, $-J_1$, $\Delta J_{\rm ext}$
and $\Delta J_H$ as given by (\ref{e:J_1_J_2_adapted})-(\ref{e:J_ext_adapted}).

Note that in the above derivation, as in the geometrical derivation of
Sec.~\ref{s:conservation_laws}, we have not assumed that the energy-momentum 
tensor $\w{T}$ obeys the spacetime symmetries. 
\end{widetext}

\end{document}